\documentclass[11pt]{article}
\usepackage{amsmath,amsthm,latexsym,amssymb,amsfonts,epsfig}

\makeatletter
\@addtoreset{equation}{section}
\makeatother

\def\be{\begin{equation}}
\def\ee{\end{equation}}
\def\ba{\begin{eqnarray}}
\def\ea{\end{eqnarray}}

\newcommand\nn{\nonumber}
\newcommand\q{\quad}
\def\Rl{{\mathchoice
{\setbox0=\hbox{$\displaystyle\rm R$}\hbox{\hbox to0pt
{\kern0.4\wd0\vrule height0.9\ht0\hss}\box0}}
{\setbox0=\hbox{$\textstyle\rm R$}\hbox{\hbox to0pt
{\kern0.4\wd0\vrule height0.9\ht0\hss}\box0}}
{\setbox0=\hbox{$\scriptstyle\rm R$}\hbox{\hbox to0pt
{\kern0.4\wd0\vrule height0.9\ht0\hss}\box0}}
{\setbox0=\hbox{$\scriptscriptstyle\rm R$}\hbox{\hbox to0pt
{\kern0.4\wd0\vrule height0.9\ht0\hss}\box0}}}}
\newcommand{\ca}{\mathcal A}

\newcommand{\cb}{\mathcal B}

\newcommand{\cm}{\mathcal M}

  \newcommand{\Fa}{\mathfrak{A}}
  \newcommand{\Fb}{\mathfrak{B}}

\newcommand{\fe}{\mathfrak{e}}

  \newcommand{\Fm}{\mathfrak{M}}

\newcommand{\so}{\mathfrak{so}}

\title{Phase space descriptions for simplicial 4d geometries}
\author{Bianca Dittrich, James P. Ryan\\
  \small Perimeter Institute, 31 Caroline St. N, Waterloo, ON N2L 2Y5, Canada.}

\begin{document}

\maketitle

\begin{abstract}
Starting from the canonical phase space for discretised (4d) BF--theory, we implement a canonical version of the simplicity constraints and construct phase spaces for simplicial geometries.
Our construction allows us to study the connection between different versions of Regge calculus 
and approaches using connection variables, such as loop quantum gravity. We find that on a fixed triangulation the (gauge invariant) phase space associated to loop quantum gravity is genuinely larger than the one for length and even area Regge calculus. Rather, it corresponds to the phase space of area--angle Regge calculus, as defined in \cite{ds} (prior to the imposition of gluing constraints, which ensure the metricity of the triangulation).
Finally, we show that for a subclass of triangulations one can construct first class Hamiltonian and Diffeomorphism constraints leading to flat 4d space--times. 
\end{abstract}

\section{Introduction}

In many approaches to quantum gravity, such as loop quantum gravity \cite{reviews}, quantum Regge calculus \cite{regge,williams} and causal dynamical triangulations \cite{cdt}, discrete models of space--time appear in one form or another. One advantage of these discrete models is that they have the potential to provide a regularisation of uv divergences \cite{hamilt, finiteness}. On the other hand, discretisation very much complicates the role of diffeomorphisms, the underlying symmetry of continuous general relativity. Moreover, as we will show in the course of the paper, even on the kinematical level, discretisation can lead to the appearance of non--metric configurations which are absent in the classical and continuum phase space description. 

These complications arise in both the path integral quantisation and canonical quantisation. Spin foam models and quantum Regge calculus involve a path integral over discretised geometries. Were diffeomorphisms to act on the configurations within these path integrals, divergences might appear, caused by an integration over the diffeomorphism gauge orbits \cite{bojo}. Thus, it is important to understand how diffeomorphisms act on discrete manifolds. In a canonical quantisation based on discretised structures, the typical problem is to obtain a consistent time evolution. In the continuum theory, time evolution is generated by the Hamiltonian and diffeomorphism constraints, which additionally generate 4d diffeomorphisms (on--shell). The group property of these diffeomorphisms is reflected in the fact that the Hamiltonian and diffeomorphism constraints are first class. In other words, in a phase space description, the Poisson brackets of these constraints vanish on the constraint hypersurface. Unfortunately, this first class property is typically lost \cite{friedman,piran, loll} if the spatial geometries are discretised. One could argue that discretisation breaks diffeomorphism symmetry \cite{gambini}, and so we do not need to care about diffeomorphisms at the discretised level. Nevertheless, diffeomorphism symmetry has to be restored in the continuum limit, as it is a symmetry of the continuum theory. Therefore, one would expect at least some approximate notion of diffeomorphism symmetry in the discretised theory. Moreover, trying to preserve some kind of diffeomorphism symmetry (that is, trying to preserve the Dirac algebra of Hamiltonian and diffeomorphism constraints) could help to resolve quantisation ambiguities \cite{alex} in the Hamiltonian constraints.

Although there is a considerable amount of work addressing these questions in different frameworks \cite{loll, galassi,hamber,david,wael,zap,bdlf3d, bdbb}, in our view the issue is not yet settled, even at the classical level. One result in this paper is that one can obtain a phase space description for $(3+1)$--dimensional discrete geometries with the following properties.  One can impose a consistent dynamics, at least on a certain class of special triangulations,  which leads to flat 4d geometries  and upon which there is a well--defined action of the diffeomorphism constraints. Since flat geometries are a subsector of the space of all solutions (of the Regge equations, for instance), we expect at least some remnant of the diffeomorphism symmetry from the full description of the theory. In fact, given a simplicial 4d manifold that satisfies the Regge equations of motion, one can always subdivide a 4--simplex of this triangulation by placing a new vertex into this 4--simplex such that the geometry of the 4--simplex is still flat. In this way, one obtains four gauge degrees of freedom, as the geometry of the triangulation is not changed and the vertex can be freely placed inside the 4--simplex. 

 Our results are similar to the work \cite{wael}  by Waelbroeck and Zapata which defines `topological gravity' as a subsector of BF--theory.  This subsector has \lq flat dynamics' and they impose conditions which allow for a geometric interpretation of the bivector field $E$ and the curvature field $F$. These conditions are a canonical version of the simplicity constraints appearing in the Plebanski formulation \cite{pleb} of general relativity. We will differ, however, in the methods and in certain parts of the conclusions. In particular, we will perform a full reduction of the BF--theory phase space to a phase space describing geometric configurations. This phase space describes `kinematical configurations', and can be taken as a starting point to define a dynamics leading to the full set of (Regge) gravity solutions.

Another motivation of this work is to explore connections among different quantum gravity approaches, mainly among spin foam models, canonical loop quantum gravity (LQG) and the different versions of Regge calculus, namely length Regge calculus \cite{tregge}, area Regge calculus \cite{barrettarea} and area--angle Regge calculus \cite{ds}. See for instance \cite{immirzi,makela} for attempts to connect the Ashtekar variables on which LQG is based, with Regge calculus. We will find a correspondence between these theories only at the kinematical (and classical) level as we do not discuss the full implementation of the dynamics here. Nevertheless, if there is a definition of the diffeomorphism and Hamiltonian constraints available in one of these versions the methods presented here allow one to translate these to the other versions.

The definition of spin foam models \cite{world} usually starts with the Plebanski action, which can be seen as a BF--theory action plus additional terms involving the simplicity constraints. These constraints impose that the bivector field appearing in this action can be written as (the Hodge dual of) a wedge product of tetrad fields. A considerable amount of current work on spin foam models \cite{epr,simet,fk,alexandrov} discusses the question of how to properly impose the simplicity constraints. This question is crucial as BF--theory in itself is a topological theory, that is, it does not have local degrees of freedom. Only if the variations of the bivector field are restricted by the simplicity constraints does one obtain local degrees of freedom. Similarly in area(--angle) Regge calculus one uses areas (and angles) as fundamental variables. These also need to be constrained to come from a consistent assignment of length variables to the edges of the triangulation, in order to obtain a theory with local degrees of freedom. 

We therefore follow \cite{wael} and start with the phase space of  discrete $SO(4)$ BF--theory. In addition to the canonical structure defined in \cite{wael}, we introduce the Barbero--Immirzi parameter following \cite{monte}, to keep in line with the starting point for many of the spin foam models (and to a lesser extent loop quantum gravity). On this phase space we will impose the so-called simplicity constraints ensuring the geometricity of the configurations, i.e. that the bivector field follows from the dual of the wedge product of vectors associated to the edges of the triangulation. The aim is to obtain a reduced phase space with respect to these constraints and hence a phase space describing simplicial 3d geometries embedded into 4d geometries. To facilitate the geometric interpretation we will work with $SO(4)$--gauge invariant quantities, that is areas, 3d and 4d dihedral angles and length variables. This also allows the connection to the different forms of Regge calculus.

 As we will see, the simplicity constraints can be subdivided into three sets which can be separately dealt with. We will obtain the phase space corresponding to loop quantum gravity restricted to a fixed triangulation\footnote{With this phase space we mean the following: we restrict the Hilbert space of loop quantum gravity to a fixed triangulation and consider the corresponding classical phase space associated to this triangulation. This phase space is different from the continuum phase space, i.e. the Ashtekar connection and conjugated densitised triad fields, upon which the LQG quantisation is based. On this continuum phase space, simplicity constraints do not appear.}
already after we have imposed the first set of these constraints. Thus, in this phase space the simplicity constraints are not fully implemented and therefore degrees of freedom are included which do not correspond to (lattice) metric degrees of freedom. The question arises of how spin foam models with proper implementation of the simplicity constraints can be connected with canonical loop quantum gravity, where the canonical version of these constraints are not implemented fully.

 Of the three subsets of simplicity constraints mentioned before, the first two sets of simplicity constraints are second class in themselves, however the last set -- giving constraints between the areas only -- is not. Here one either reduces by these first class constraints or introduces further constraints conjugated to these area constraints. These further constraints might arise as secondary constraints from the primary simplicity constraints, since in the continuum \cite{canpleb} there are secondary constraints coming from the Poisson brackets of the simplicity constraints with the Hamiltonian and diffeomorphism constraints. 

Such left--over simplicity constraints do not appear in (2+1)--dimensional gravity. Thus, also the phase space associated to loop quantum gravity restricted to a fixed triangulation corresponds exactly to the one from length Regge calculus \cite{bdlf}.  The reason  why the issue of defining a phase space for $(3+1)$ simplicial triangulations is more involved than in $(2+1)$ dimensions is the following: the natural configuration variables for a simplicial manifold are the length variables associated to the edges of the triangulation. These length variables specify completely the internal geometry of the triangulation and are moreover unconstrained (apart from the triangle inequalities and higher dimensional generalisations which we will ignore). The natural variables to encode the extrinsic curvature are the 4d dihedral angles between neighbouring tetrahedra and therefore based on the triangles of the triangulation. Now in general for non--degenerate simplicial 3d (and 4d) manifolds the number of triangles is greater than the number of edges. The problem of left--over simplicity constraints is therefore rooted in the discretisation by triangulation, which leads to a mismatch in the number of variables, which are, however, in one-to-one correspondence in the continuum.

The structure of the paper is as follows.  In Section \ref{cans}, we shall define the canonical variables and Poisson structures for a discrete BF--theory.  In the subsequent section, we shall develop a complete set of simplicity constraints for the discrete phase space.  We prove that these constraints are sufficient to ensure the geometricity of the 3d triangulation in Section \ref{proof}. We introduce gauge--invariant variables in Section \ref{gauge}, which will provide the necessary parameters for the Gau\ss~reduced phase space.  We consider the explicit case of the boundary of a 4-simplex, and through Section \ref{simplex} we enforce the Gau\ss~and simplicity constraints to arrive at the reduced phase space of geometric configurations. In particular, we divide the examination of the simplicity constraints into two subsections.  The first, subsection \ref{lr}, deals with the constraints implementing the equality of left-- and right--handed sectors.  The second, subsection \ref{glue}, reduces by the constraints ensuring that the edge lengths are independent of the tetrahedron in which they are calculated.   In Section \ref{general}, we expand our analysis to more general configurations, while in the penultimate section, we consider implementing the dynamical constraints on the phase space.  Finally, we recapitulate and conclude in Section \ref{conclude}.



\section{Canonical structures for 4D BF--theory}
\label{cans}

In this section we will collect some basic material and notation on the canonical formulation of discretised BF theory as this will provide us with the phase space on which we will implement the simplicity constraints. Additional material can be found in Appendices \ref{def} and \ref{discrete}. 

Canonical structures for continuum BF--theory are dealt with extensively in \cite{monte}.  We shall develop analogous structures for the discrete setting.  The action for 4d BF-theory with an extra \lq topological'  term is
\be\label{can1}
	{\cal S}_\cm[E,A]  	=  \int_\cm \left(E+\frac{1}{\gamma} {}^* E\right)\wedge F
 				=\int_\cm E\wedge \left(F+\frac{1}{\gamma}{}^* F\right),
\ee
where ${\cm}$ is a smooth manifold,   $E$ is an $\so(4)$-valued 2-form and $A$ is the $\so(4)$-valued 1-form. They are referred to as the bivector and connection fields, respectively.  The curvature of this connection is $F = dA + A\wedge A$.  We shall refer to the coupling constant $\gamma$, as the Barbero-Immirzi parameter even in the non-gravitational theory.   

It is straightforward to see that the connection $A$ is canonically conjugate to the field $\Pi:=E+\frac{1}{\gamma}{}^\star E$. In fact, we shall wish to deal mainly with the bivector field which we may express in terms of the canonical momentum as
\be\label{can2}
	E=\frac{\gamma^2}{\gamma^2-1}\left(\Pi-\frac{1}{\gamma}{}^*\Pi\right).
\ee
As is well known, the $\so(4)$ Lie algebra splits into selfdual and anti-selfdual sectors. We may make this distinction explicit by introducing selfdual and anti-selfdual fields $E_{\pm} = \frac{1}{2}(E \pm {}^*E)$, $A_\pm = \frac{1}{2}(A\pm{}^*A)$,\footnote{This implies $F_{\pm} = F[A_\pm]$.} which results in
\be
	{\cal S}[E_{\pm},A_{\pm}] = \left(1+\frac{1}{\gamma}\right)\int E_+ \wedge F_+  + \left(1-\frac{1}{\gamma}\right)\int E_- \wedge F_-\;.
\ee
Note, the special case $\gamma=\pm 1$. In this scenario, the action contains only the selfdual fields or the anti-selfdual fields, respectively. Thus, only half the number of variables are present in comparison to the case of general $\gamma$. In this case, the (anti-)selfdual connection $A_\pm$ is directly conjugate to the (anti-)selfdual field $E_\pm$.

Now, we shall repeat this process in the discrete setting. We describe this undertaking in Appendix \ref{discrete},  but we shall reiterate some of the pertinent details here.   We pass to the Hamiltonian formalism and replace the continuum 3d hypersurface by a 3d simplicial complex. Likewise, we define fields on sub--simplices of  this discrete manifold.    The discrete analogue of the action is
\be\label{bf1}
	{\cal S}[\Pi_{ij}, M_{ij},N_\fe, \tilde{N}_i] = \int_{\mathbb{R}}dt\left[\sum_{\{ij\}} C^{ABC} \Pi_{ij}^A [M_{ij}^{-1}]^{BD}\dot{M}_{ij}^{DC} + \sum_{e}N_e^A F_e^A + \sum_{i}\tilde{N}_i^A G_i^A\right](t).
\ee
where we label the tetrahedra by the index $i$, so that one can pick out an oriented triangle $\{ij\}$ by the pair of tetrahedra that share it.   Moreover, $e$ denotes an edge of the simplicial complex.
The canonical variables are associated to the triangles $\{ij\}$ where $(\Pi_{ij}, M_{ij})$ are the discrete counterparts of the canonical variables $(\Pi, A)$.  $M_{ij}$ is a rotation between the frames of reference in the adjacent tetrahedra,  and $\Pi_{ij} = E_{ij}+\frac{1}{\gamma}{}^\star E_{ij}$.  The latter two terms in (\ref{bf1}) are the flatness and Gau\ss~constraints.  We discuss their significance later on, but they are enforced by Lagrange multipliers $N_e$ and $\tilde {N}_i$, respectively.   The Poisson brackets between the bivectors $E_{ij}$ and the holonomies $M_{ij}$ are then given by
\ba\label{can3}
 \{E^A_{ij},M_{kl}^{BC}\} & = & \frac{\gamma^2}{\gamma^2-1}\,\left(\delta^{AA'}-\frac{1}{\gamma}\epsilon^{AA'}\right)\left(\delta_{ik}\delta_{jl}\,C^{A'BD}M_{kl}^{DC} + \delta_{il}\delta_{jk} \,C^{A'CD}M_{kl}^{BD}\right),   \\
\{E^A_{ij},E^B_{kl}\} & = & \delta_{ik}\delta_{jl} \frac{\gamma^2}{\gamma^2-1}C^{ABC}\left(\delta^{CC'}-\frac{1}{\gamma} \epsilon^{CC'}\right)E_{kl}^{C'}\;.
\ea

Between the canonical variables we have two kinds of relations which are consistent with the Poisson brackets (\ref{can3}). Firstly, $M_{ji}$ is the rotation inverse to $M_{ij}$ and secondly, we have the relation $E_{ji}^A=-M_{ij}^{BA}E_{ij}^B$ which relates the bivector as seen in both tetrahedra adjacent to the triangle $\{ij\}$.

In this setting, we may also perform the decomposition into selfdual and anti-selfdual parts.   The projected fields and structure constants are $E_{\pm ij}^A$, $M_{\pm ij}^{AB}$  and $C_{\pm}^{ABC}$ respectively, which we detail in Appendix \ref{def}.  In summary, we obtain the following symplectic structures (assuming $i = k$ and $j=l$)
\ba\label{can7}
 	\{E^A_{\pm},M_{\pm}^{BC}\}    &=&\frac{\gamma}{\gamma\pm 1}C^{ABD}_\pm M_{\pm}^{DC}\;,  \nn\\
	\{E^A_\pm,E^B_\pm \}  &=&\frac{\gamma}{\gamma\pm 1}C_\pm^{ABC}E^C_{\pm}\;.
\ea
The constraints in terms of $(E_{ij}, M_{ij})$ are the discrete flatness and Gau\ss~constraints
\ba\label{flata}		
F_e^A & = & 
C^{ABC} \left(M_{ij}M_{jk}\dots M_{ni}\right)^{BC}\;,\\
	\label{close}	G_i^A & = & \sum_{k}E_{ik}^A\;.
\ea
where the product on the right hand side of (\ref{flata}) is the holonomy around the edge $e\in\Delta$, and the sum on the right hand side of (\ref{close}) imposes closure of the four bivectors associated to the tetrahedron $\{i\}$. Both Gau\ss~ constraints and flatness constraints form a first class constraint set. The Gau\ss~constraints generate $SO(4)$ gauge transformations
\ba
M_{ij}^{AB} & \rightarrow &  \Lambda_{i}^{AA'} M_{ij}^{A'B'}(\Lambda^{-1})_{j}^{B'B}\;,
\nn\\
E_{ij}^A      &\rightarrow& \Lambda_{i}^{AA'}E^A_{ij}\;,
\ea
for $SO(4)$ matrices $\Lambda_{i},\Lambda_{j}$ associated to the tetrahedra $\{i\}$ and $\{j\}$ respectively. The flatness constraints generate the translation symmetry 
\ba
M_{ij}^{AB} &\rightarrow& M_{ij}^{AB} \;,\nn\\
E_{ij}^{A} &\rightarrow& E_{ij}^A + \sum_{e @ \{ij\}} \Lambda_{ei}^A\;,
\ea
where the $\Lambda_{ei}$ are gauge parameters associated to the edges of the triangulation.  They are expressed in the reference system of the tetrahedron $\{i\}$ and we sum over the three edges of the triangle $\{ij\}$, see Appendix \ref{flat} for more details on the action of the flatness constraints.

\section{Canonical simplicity constraints}\label{simplicity}

In this section, we will introduce a canonical version of the simplicity constraints, which will impose the metricity of the bivector fields. That is, for phase space points satisfying these constraints, the bivectors can be written as (the Hodge duals of) wedge products of vectors that can be consistently associated to the edges of the triangulation.

The action (\ref{can1}) describes a topological field theory, namely $BF$--theory. Adding to this action another term, known as the simplicity term
\be\label{final1}
{\cal S}[E,\phi]_{simp} = -\frac{1}{2}\int \phi^{AB} E^A \wedge E^B\;,
\ee
one arrives at an action which describes gravity (and another topological theory) \cite{pleb,world}. $\phi^{AB} = \phi^{BA}$ is the Lagrange multiplier which ensures that the bivector field arises as $E = \pm^* e\wedge e$ or $E = \pm e\wedge e$.  The first pair of solutions are known to give rise to the gravitational sector of the theory, while the second pair give rise to the topological sector.

As one might expect, we wish to have a comparable set of simplicity conditions for our discrete theory. But before we proceed, let us note that this action could be discretised using a 4d triangulation. But to perform a canonical analysis of the discretised action is in general quite difficult. A canonical analysis of the continuum action has only appeared recently \cite{alexandrov}. One of the problems that a discrete approach faces is to identify spatial slices in a general 4d triangulation. Generically, the number of variables associated to different spatial slices will vary. One could remedy this situation by considering a special triangulation, for instance a triangulation that fits into a hypercubical lattice. One could then restrict the class of spatial slices to those which correspond to \lq planes' in this hypercubical lattice. This would, however, introduce a preferred slicing.   

As can be seen in Section \ref{cans}, we will follow a different route here and discretise only the spatial hypersurface.  In fact, we impose constraints on the canonical phase space of discretised $BF$--theory that ensure the geometricity of the configuration. The aim will be to reduce the phase space with respect to these constraints and the Gau\ss~constraint generating $SO(4)$ gauge transformations. The phase space obtained in this manner should correspond to the one describing true geometric triangulations. As the conditions for such a phase space, we will ask that every phase space point allows one to determine (at least locally) a set of consistent edge-lengths for the associated triangulation. Furthermore, we require a non--degenerate symplectic form on this phase space.  It is on this reduced phase space that one could analyse the effect of instituting the diffeomorphism and Hamiltonian constraints.

We will start to discuss such geometricity conditions for a general 3d triangulation. These constraints will be highly redundant, however, and therefore it is difficult to obtain the reduced phase space in the general case. As a result, it will turn out to be enlightening to first consider a simple closed 3d triangulation, namely the 3d boundary of a 4--simplex.  Later on, we will consider more complicated triangulations.



As a first set of constraints, for every tetrahedron we have the discrete equivalent of the Gau\ss~constraints, ensuring that the tetrahedra are closed
\be
\sum_{j}E^A_{ij}=0\;,
\ee
where the sum is over the four tetrahedra $\{j\}$ adjacent to the tetrahedron $\{i\}$.  Furthermore, we will introduce the well known diagonal simplicity and cross simplicity constraints
\ba\label{geo3}
S_{ij}:=\epsilon^{AB}E^A_{ij}E^B_{ij}=0\;  , \q\q
S_{ijk}:=\epsilon^{AB}E^A_{ij}E^B_{ik}=0\;, 
\ea
among the bivectors belonging to one tetrahedron $\{i\}$. (These constraints have been used also in \cite{wael}.) The diagonal simplicity and cross simplicity constraints ensure that every bivector $E^A$ spans only a two--dimensional subspace and that every pair of bivectors from one tetrahedron spans only a three--dimensional subspace.

We can reformulate  the simplicity and cross simplicity conditions (\ref{geo3}) using the splitting of the fields into selfdual and anti--selfdual terms. We define
\be\label{geo4}
A_{ij\pm}=E_{ij\pm} \cdot E_{ij\pm}\;,\q\q A_{ijk_\pm}=E_{ij\pm}\cdot E_{ik\pm}\;,
\ee
where $T_1\cdot T_2= T_1^{A}T_2^{A}$.  Then, we can rewrite the constraints as
\ba\label{geo4b}
S_{ij}=A_{ij+}-A_{ij-}\; ,\q\q S_{ijk}=A_{ijk+}-A_{ijk-}\;,
\ea
that is the 3--geometry as defined in the selfdual sector coincides with the three--geometry as defined in the anti--selfdual sector, see also \cite{reisenberger1}.

These conditions are not sufficient. We have also to ensure that all the bivectors meeting at an edge span only a three--dimensional space, as every bivector should be normal to the edge in common. Therefore, we add the edge simplicity constraints\footnote{This set of constraints can be replaced by an alternative set proposed in \cite{wael}. As discussed in Appendix \ref{appalt} this alternative set distinguishes between the gravitational and the topological sector, in contrast to (\ref{geo5}).}
\be\label{geo5} 
C_{ijkl}=\epsilon^{AB}\, E_{ik}^A \, M_{ij}^{BC}E_{jl}^C   \; ,
\ee
where $E_{ik}^A$ and $E_{jl}^B$ are bivectors associated to two triangles meeting at an edge and belonging to neighbouring tetrahedra $\{i\}$ and $\{j\}$. Thus, we do not include all pairings of triangles meeting at an edge. If more than five tetrahedra share an edge, it might happen, that two triangles meet in an edge which are not from neighbouring tetrahedra. Nevertheless, (by transitivity) the constraints (\ref{geo4}) are sufficient. Note that the constraints (\ref{geo5}) involve -- in contrast to the covariant simplicity constraints \eqref{final1}-- the connection variables. As we will see later on this set of constraints implies the gluing conditions \eqref{53}, which are not realised in a phase space associated to loop quantum gravity restricted to a fixed triangulation.

\section{Simplicity constraints ensure geometricity}
\label{proof}

Here we will show that the simplicity, cross simplicity and edge simplicity constraints ensure that the bivectors come from (duals of) wedge products of vectors associated to the edges of the triangulation.

Note that the diagonal simplicity, cross simplicity and edge simplicity constraints are left invariant under the transformation $E \rightarrow {}^*E$ of the bivectors, which exchanges the gravitational sector with the topological sector. Hence these constraints do not distinguish between the sectors; see, however, the discussion in Appendix \ref{appalt}.

Consider two bivectors, $E_{ij}$ and $E_{ik}$, associated to the tetrahedron $\{i\}$. The diagonal simplicity and cross simplicity constraints ensure \cite{puzio} that the bivectors are simple and have a common factor, that is
\be\label{simp1}
E_{ij}=u\wedge v \;, \q E_{ik}= u \wedge w   . 
\ee
For three simple bivectors $E_{ij},E_{ik}$ and  $E_{il}$ there are two possibilities to ensure that every pair spans only a two--dimensional subspace, namely
\be\label{simp2}
\begin{array}{lclclcl}
{}^* E_{ij} &= &c_j\, s \wedge t \; ,    			&&						{}^* E_{ij} &=& x_{ij} \wedge n_i \;, \\
{}^* E_{ik}&= &c_k \, t \wedge u \;,     			&\q\q\q\q\text{or}\q\q\q\q&	   			{}^* E_{ik} &=& x_{ik} \wedge n_i \;, \\
{}^* E_{il}&=  &c_l \, u \wedge s  \;, 			&&						{}^* E_{il} &=& x_{il} \wedge n_i \;,
\end{array}
\ee
where $s,t,u$ and $x_{ij},x_{ik}, x_{il},n_{i}$ are vectors while $c_j,c_k$ and $c_l$ are scalars.

Note that the two solutions can be mapped to each other by applying the Hodge dual to the bivectors.  
The right hand side of (\ref{simp2}) leads to the so--called topological sector, where the dual bivectors are equal to the wedge products between the normal to the tetrahedron, $n_i$, and the normals to the triangles $x_{ij}, x_{ik}, x_{il}$.  The left hand side corresponds to the gravitational sector where the dual bivectors are given by wedge products of edge vectors. 

On the gravitational sector, the dual of the fourth bivector $E_{im}$ associated to the tetrahedron $\{i\}$ satisfying diagonal simplicity and cross simplicity with the other three bivectors can be written in the form
\ba\label{simp3}
{}^* E_{im}&=&(u+ \alpha s ) \wedge ( \beta s +\delta t), \nn\\
&=& \beta\, u\wedge s -\delta\, t\wedge u +\alpha\delta\, s\wedge t   \; .
\ea
By rescaling $u$ and $t$ in (\ref{simp2}) we can assume w.l.o.g.  that in (\ref{simp2}) $c_l=1$ and $c_k=1$. Multiplying the Gau\ss~constraint $\sum_{n} {}^* E_{in}=0=\sum_n E_{in}$ with $n_{ij}\wedge n_{ik}$,\,\,$n_{ik}\wedge n_{il}$ and $n_{il}\wedge n_{ij}$ respectively (where $n_{ij}$ is normal to $s$ and $t$ and so on,) we can conclude that
\be\label{simp4}
\beta=-1\;,\q\q \delta=1 \; ,\q\q  c_j=-\alpha \; .
\ee 
Finally, we can absorb the factors of $\alpha$ into a redefinition of the one--vectors $s':=|\alpha|^{1/2}s,\, t':=|\alpha|^{1/2}t,\, u':=|\alpha|^{-1/2}u$, where in the following we will drop the prime. In summary, we obtain for the gravitational sector
\be
\begin{array}{lcl}
\label{simp5}
{}^*E_{ij}&=&\pm s\wedge t\;,\\
{}^*E_{ik}&=& t\wedge u\;,\\
{}^*E_{il}&=& u\wedge s\;,\\
{}^*E_{im} &=& (u\mp s)\wedge(t-s)  .
\end{array}
\ee
We shall denote by $v_{im}$ the vertex of the tetrahedron $\{i\}$ which is opposite the triangle $\{im\}$. If $(s,t,u)$ is a triple of positively oriented vectors starting from $v_{im}$, then,  with the upper choice of sign, (\ref{simp5}) defines the inward pointing bivectors corresponding to the triangles.  If we reverse the direction of $u$ in the triple $(s,t,u)$, then (\ref{simp5}) with the lower choice of sign defines outward pointing bivectors. Redefining $u'=-u$ for this case, so that $(s,t,u')$ is again a positive oriented triple of edges starting from $v_{im}$, gives ${}^*E_{ij}=-s\wedge t$,\,\,${}^*E_{ik}=-t\wedge u'$ and so on.

Hence, we obtain the well known result \cite{puzio} that the diagonal simplicity, cross simplicity and Gau\ss~constraints allow for four different types of solutions.   For the gravitational sector, either the dual bivectors or minus the dual bivectors are coming from a tetrahedron. Since applying the Hodge dual to the bivectors preserves the constraints, we have also two other types of solution, namely that the bivectors or minus the bivectors are coming from a tetrahedron.

So far we obtained this result for each tetrahedron separately. But we must also ensure that these tetrahedra glue to each other consistently. For simplification, we will restrict our considerations to the gravitational sector.

Consider two neighbouring tetrahedra $\{i\}$ and $\{j\}$ and three bivectors $E_{ij}$, $E_{ik}$, $M_{ij}E_{jk'}$ meeting at an edge. Because of the previous discussion, we can conclude that 
\begin{displaymath}
\label{simp6}
\begin{array}{lclclcl}
{}^*E_{ij} &=&\sigma s\wedge t  \; , &\q\q\q\q&{}^* M_{ij}E_{ji}&=&\sigma' s'\wedge t' \;,~ \nn\\
{}^* E_{ik}&=&\sigma t\wedge u \; ,  &&{}^* M_{ij}E_{jk'}&=&\sigma' u'\wedge t'  \;,~
\end{array}
\end{displaymath}
where $(s,t,u)$ is a positively oriented triple, $(s',t',u')$ is negatively oriented and $\sigma,\sigma'=\pm$ encode the sign ambiguity.

From the condition (\ref{tri}), we have $s\wedge t \sim  s'\wedge t'$  and hence both $s'$ and $t'$ are linear combinations of $s$ and $t$. 
The edge simplicity constraint (\ref{geo5}) ensures that the four vectors $t,t'$ and $u,u'$ span a three--dimensional subspace. If $u'$ is a linear combination of $s,t$ and $u$ all the simplicity relations are satisfied. In this case, however, the two tetrahedra span the same three--dimensional subspace, that is the dihedral angle between them is vanishing. Therefore let us assume the generic case, that is that the inner product between $u'$ and the normal to the tetrahedron $\{i\}$ is not vanishing. Since $t',u',t,u$ span a three--dimensional subspace we can conclude that in this case $t'$ is a linear combination of $t,u$ and $u'$. At the same time, $t'$ is a linear combination of $s$ and $t$,
\be\label{simp7}
t'=\alpha t+\beta u +\delta u'=\alpha' s + \beta' t \; .
\ee
Taking the inner product with a vector normal to $t,u$ and $u'$ we obtain $\alpha'=0$ and hence $t'=\alpha t$. In the same way, we can conclude from the edge simplicity at $s$ that $s'=\beta s$ and using $\sigma s\wedge t =-\sigma'  s'\wedge t'$ we have $\alpha= -\sigma\sigma' \beta^{-1}$.  
So far, we obtained
\begin{equation}\label{simp6b}
\begin{array}{lclclcl}
{}^* E_{ij}&=&\sigma s\wedge t \;,  &\q\q\q&{}^* M_{ij}E_{ji}&=&-\sigma s\wedge t\; ,\\
{}^* E_{ik}&=&\sigma t\wedge u \;,  &&{}^* M_{ij}E_{jk'}&=& \alpha \sigma' t\wedge u'  \;,\\
{}^* E_{il}&=&\sigma u\wedge s  \;,  &&{}^* M_{ij}E_{jl'}&=& -\alpha^{-1}\sigma u'\wedge s  \;,\\
{}^* E_{im}&=&\sigma (u-s)\wedge (t-s)\;, &&{}^* M_{ij}E_{jm'}&=& \sigma'(u' +\alpha^{-1}\sigma\sigma'  s)\wedge (\alpha t+\alpha^{-1}\sigma\sigma' s) \; .
\end{array}
 \end{equation}
From the (edge) simplicity between ${}^*E_{im}$ and ${}^*M_{ij}E_{jm'}$ we have
\be
\alpha t +\alpha^{-1} \sigma\sigma' s = \lambda (u-s) +\mu  (t-s) + \nu (u'+\alpha^{-1}\sigma\sigma'  s) \; .
\ee
Again, by taking the inner product with the vector normal to the tetrahedron $\{i\}$, we can conclude that $\nu=0$. From the inner product with the normal to $\{j\}$ we have that $\lambda=0$. Hence, $\alpha t+\alpha^{-1}\sigma\sigma' s= \mu t -\mu s$, from which we can conclude that $\sigma\sigma'=-1$ and $\mu=\alpha=\alpha^{-1}=1$. Hence, in the generic case of non--parallel tetrahedra we have proven that the dual bivectors can be consistently written as wedge products of edge vectors also for neighbouring tetrahedra. Moreover the choice of orientation is the same, i.e. the bivectors are either inward or outward pointing for pairs of neighbouring tetrahedra.

\section{ Gauge invariant quantities}
\label{gauge}

In this section we will construct a complete set of $SO(4)$ gauge invariant quantities. These will simplify the implementation of the simplicity constraints and moreover make the connection to the different version of Regge calculus obvious.

Gauge invariance is not broken by projecting down onto the selfdual and anti-selfdual sectors.  Thus, we shall find it convenient to define gauge invariant quantities for both the left-handed and right-handed sectors in parallel.   The easiest gauge invariant quantities to construct are the squared areas $A_{ij\pm}$ and the 3d dihedral angles
\be\label{cphi}
\cos \phi_{ijk\pm} := \frac{A_{ijk\pm}}{\sqrt{A_{ij\pm}A_{ik\pm}}}= \frac{E_{ij\pm}\cdot E_{ik\pm}}{\sqrt{E_{ij\pm}\cdot E_{ij\pm}\,\, E_{ik\pm}\cdot E_{ik\pm}}}  \; .
\ee
 As already noted, the diagonal simplicity and cross simplicity constraints imply that the areas and the 3d dihedral angles computed from the left-- or right--handed sectors coincide.

Another class of variables is the 4d dihedral angles.  We define bi-vector \lq normals' as $N_{ijk\pm}^A := (E_{ij\pm}\times E_{ik\pm})^A = \pm C^{ABD}_\pm E_{ij\pm}^B E_{ik\pm}^C$ and the angle between two of them as\footnote{This is in close analogy with the state of affairs in 3d space-time, where one defines the normal to the triangle to be $n^{a} = \epsilon^{abc} e_i^b e_j^c$ and $e^a$ are the triad vectors.}
\be\label{cons2}
\cos\theta_{ik,jl\pm} := \frac{N_{ijk\pm}^B \, M^{BC}_{ij\pm} N^C_{jil\pm}}{\sqrt{N_{ijk\pm}\cdot N_{ijk\pm}\,\, N_{jil\pm}\cdot N_{jil\pm} }} \; .
\ee
where the result may not only depend on the choice of tetrahedra $\{i\}$ and $\{j\}$ but a priori also on the choice of triangles $\{ik\}$ and $\{jl\}$.

As we will see, this is not the situation in the geometric sector.  For the case that the three triangles in question share the same edge, we can write
\begin{xalignat}{2}
&{}^* E^{aa'}_{ij}=2e_1^{[a}e_2^{a']} \;,  && {}^* M_{ij}E_{ji}^{aa'}=-2e_1^{[a}e_2^{a']} \;,\nn\\
&{}^* E^{aa'}_{ik}=2e_3^{[a}e_1^{a']}\;,   && {}^* M_{ij}E_{jl}^{aa'}=-2{e'}_3^{[a}e_1^{a']}\;,
\end{xalignat}
where $(e_1,e_2,e_3)$ is a positively (negatively) oriented triple of vectors whereas $(e_1,e_2,e'_3)$ is a negatively (positively) oriented triple. For the normal bivectors $N$, we obtain
\begin{xalignat}{2}
&N_{ijk}^{aa'}=2e_1^{[a}n_i^{a']} \;,&& N_{jil}^{aa'}=2e_1^{[a}n_j^{a']}\;,
\end{xalignat}
where $n_i^a=\epsilon^{abcd} e_1^b e_2^c e_3^d$ is the positively (negatively) oriented normal to the tetrahedron $\{i\}$ and $n_j^a=\epsilon^{abcd} e_1^b e_2^c {e'}_3^d$ is the negatively (positively) oriented normal to the tetrahedron $(j)$. Thus, $\cos\theta_{ij,kl}$ gives the (inner) dihedral angle between the two normals to the two tetrahedra in the case that $\{ij\},\{ik\}$ and $\{jl\}$ have an edge in common. Also note that in this case, $\theta_{ik,jl}$ does not depend on the choice of the triangles $E_{ik},E_{jl}$ as long as they share the same edge with $E_{ij}$.

To simplify the expression (\ref{cons2}) for the 4d dihedral angles, we need the formula (\ref{cons1}) giving the contraction of two structure constants. Applying this to (\ref{cons2}), we get
\be
\cos \theta_{ik,jl\pm } =- \frac{ A_{ij\pm}\,\,  E_{ik\pm}\cdot M_{ij\pm}E_{jl\pm} + A_{ijk\pm} A_{jil\pm}   }   {  \sqrt{ A_{ij\pm} A_{ik\pm} -A_{ijk\pm}^2   } \,\,  \sqrt{A_{ji\pm} A_{jl\pm} -A_{jil\pm}^2 }}
=- \frac{\cos \rho_{ik,jl\pm} + \cos \phi_{ijk\pm}\cos\phi_{jil\pm}}{\sin\phi_{ijk\pm}\sin\phi_{jil\pm}}  \;,
\ee
where we introduced 
\be
\cos \rho_{ik,jl\pm}=\frac{  E_{ik\pm}\cdot M_{ij\pm}E_{jl\pm}   }
                          { \sqrt{A_{ik\pm}A_{jl\pm}}} \; .
\ee
If the three triangles $\{ij\},\{ik\}$ and $\{jl\}$ share an edge we obtain from the edge simplicity constraint that $\cos \rho_{ik,jl+}=\cos\rho_{ik,jl-}$ and hence we have that the 4d dihedral angles are equal up to a sign and  $\cos \theta_{ik,jl+}=\cos \theta_{ik,jl-}$ on the geometric (and topological) sector.

\section{The phase space for a 4--simplex}
\label{simplex}

Unfortunately, the system of simplicity and Gau\ss~constraints is, in general, quite redundant. This complicates the determination of the Dirac brackets. To obtain an irreducible set of constraints, we will therefore consider a very simple triangulation, namely the boundary of a 4--simplex, consisting of five tetrahedra, where every tetrahedron is glued to the other four tetrahedra. The first part of the reduction process, namely the one which says that the left--handed geometry coincides with the right--handed geometry, carries over to general triangulations. But we will see, that we have left--over constraints, which we shall implement later on.

The (3d) boundary triangulation of a 4--simplex has $10$ triangles $\{ij\},i<j$ and $5$ tetrahedra $\{i\},i=1,\dots,5$. We have therefore $6\times 10$ configuration variables and $6\times 10$ momentum variables. Since we have $6$ constraints per tetrahedron due to the closure constraints, we obtain $30$ (first class) constraints. (Although one might first think that there is some redundancy in the closure constraints, this is not the case.  Redundancy occurs only in the case that the momentum variables $M^{AB}_{ij}$ are trivial.) Therefore,  we expect $60$ phase space variables in the Gau\ss--reduced sector. 

These $60$ variables can be chosen from the following set of 140 variables
\ba\label{41}
A_{ij\pm}&:=&E_{ij\pm}\cdot E_{ij\pm} \;,\nn\\
A_{ijk\pm}&:=&E_{ij\pm}\cdot E_{ik\pm}\;,\nn\\
\cos\theta_{ijl\pm}&:=&\cos\theta_{il,jl\pm}=\frac{N_{ijl\pm}^B \, M^{BC}_{ij\pm} N^C_{jil\pm}}{\sqrt{N_{ijl\pm}\cdot N_{ijl\pm}\,\, N_{jik\pm}\cdot N_{jil\pm} }}  \; .
\ea
The final set of variables may be somewhat puzzling at first sight, but we should notice that they are an artefact of the connectivity of a 4-simplex.  Each edge is contained in exactly  three triangles and three tetrahedra.

The $20$ variables $A_{ij\pm}$ do not depend on whether they are computed in the tetrahedron $\{i\}$ or the tetrahedron $\{j\}$, that is $A_{ij\pm}=A_{ji\pm}$. A priori, we have $60$ variables $A_{ijk\pm}$, namely $2\times 6$ per tetrahedron. We can use the Gau\ss~constraints for a tetrahedron to reduce the number of independent variables to the $2\times 2$ non--opposite angles $\phi_{ijk\pm}$.\footnote{$\cos\phi_{ijk\pm}$ is given in (\ref{cphi})} Here, we have to choose some prescription within the tetrahedron $\{i\}$ for $j_I,k_I$ with $I=1,2$ , such that the resulting angles $\phi_{ij_Ik_I}$ are non--opposite. To extract the other $2\times 4$ angles in the tetrahedron $\{i\}$,  one has to multiply the Gau\ss~constraints for this tetrahedron with any of the $E$--bivectors. In other words, one uses the $2\times 4$ equations
\be\label{43}
\sum_{k \neq i} E_{ij\pm}\cdot E_{ik\pm}=  \sum_{k\neq i} A_{ijk\pm}=   0  \; .
\ee 
At this stage, we have  $20$ area--variables and $2\times 2 \times 5=20$ angle variables. These uniquely determine the left-- and right--handed intrinsic geometry of the five tetrahedra. (Remember that the geometry of a tetrahedron is determined by six quantities, for instance the length of the edges.) Note that these variables do not just encode the information about the bivectors $E$, but also part of the information encoded in the holonomies $M$. This happens because of the condition $E_{ij}=-M_{ij}E_{ji}$ relating the bivectors in neighbouring tetrahedra by a holonomy.

We are left to choose  $20$ variables for the extrinsic geometry, that is $2$ variables per triangle. We therefore choose one of the three $\cos\theta_{ijl\pm}$ per triangle $\{ij\}$ and per $\pm$--sector. This does not necessarily mean that the angles $\theta_{ijk\pm}$ are the same (yet) for different choices of the edge in the triangle to which they are attached, just that the three angles per sector per triangle are related to each other by relations involving the other variables. 

As we have seen, from the simplicity, cross simplicity and edge simplicity constraints we can conclude that the geometric quantities as computed from the left--handed sector should coincide with the ones from the right--handed sector. That is, we have the $30$ constraints
\ba\label{44}
S_{ij} &=&A_{ij+}-A_{ij-}  \;,\nn\\
S_{ij_Ik_I} &=&A_{ij_Ik_I+}-A_{ij_Ik_I-} \;,\nn\\
C_{ijl} &=&\cos\theta_{ijl+}-\cos\theta_{ijl-} \;, 
\ea
with the understanding that in the last line of (\ref{44}), the index $l$ depends on the other two indices $i,j$. 

Note that we did not take into account all the edge simplicity constraints yet.  This can be seen from a simple counting argument.  We know that the Gau\ss~ and simplicity constraints ensure the geometricity of the triangulation.  Thus,  the reduced phase space of this triangulation should be parametrised by 20 variables, for example, the edge lengths and their conjugate momenta.  But after imposing the equality of the left--handed and  right--handed sectors (\ref{44}), we are still left with 30 variables.  Therefore, we must find some more constraints to further reduce this phase space.

Already at this stage, however,  the set of constraints (\ref{44}) is a second class system with an invertible Dirac matrix. The reason for this is that the matrix of Poisson brackets between these constraints has a triangular block structure, which allows one to compute with ease the determinant of this matrix. This will also hold for a general 3d triangulation.   With this in mind, we shall reduce our phase space in two stages.  Initially, we will compute the Dirac bracket of the first subset of simplicity constraints (\ref{44}), and reduce the phase space accordingly.  Only afterwards will we introduce another set of constraints, argue that they are necessary to capture the geometric sector and reduce once again.

\subsection{Reduction by the first subset of simplicity constraints}
\label{lr}

Here we want to compute the Poisson brackets between the subset of simplicity constraints (\ref{44}) in order to determine the Dirac bracket. As the Dirac bracket of an arbitrary phase space function with the constraints (\ref{44}) vanishes, we expect that in the reduced phase space, the sectors are set to be equal and that the flow generated by quantities from the right--handed sector coincides with the flow generated from the corresponding left--handed quantities. That is, right--handed and left--handed quantities cease to commute.

The brackets among the simplicity constraints $S_{ij}$ and between the simplicity and cross--simplicity constraints vanish, as the squared areas commute with any function of the bivectors $E$ alone. The only non--vanishing bracket involving the simplicity constraint $S_{ij}$ is with $C_{ijl}$, that is,  this block in the matrix of Poisson brackets is also diagonal. More precisely, we have
\ba\label{45}
	\{S_{ij},C_{ijl}\}    =        2\frac{\gamma^2}{\gamma^2-1}\left((\Sigma_{ijl+}-\Sigma_{ijl-})  -       \frac{1}{\gamma} (\Sigma_{ijl+}+\Sigma_{ijl-})      \right)  \; ,
\ea
where
\be\label{46}
	\Sigma_{ijl\pm}          =  \frac{E_{ij\pm}\cdot( N_{ijl\pm} \times (M_{ij\pm}N_{jil\pm}))}         {\sqrt{N_{ijk\pm}\cdot N_{ijk\pm}\,\, N_{jik\pm}\cdot N_{jik\pm}}}\;,
\ee
as calculated in more detail in Appendix \ref{form}.
Note that the first summand $(\Sigma_{ijl+}+\Sigma_{ijl-})$ vanishes on the constraint hypersurface due to equation (\ref{sun5a}).


Next we will consider the brackets among the cross simplicity constraints. Here the cross--simplicity constraints associated to different tetrahedra commute, so that the only non--vanishing brackets are between the pairs of constraints from the same triangle. That is the matrix-block associated to the cross--simplicity constraints consists of $5$ antisymmetric blocks of size $2\times 2$ on the diagonal. We have
\be\label{47}
\{S_{ijk},S_{ijl}\}=\frac{\gamma^2}{\gamma^2-1} \left((V_{ijkl+}-V_{ijkl-}) - \frac{1}{\gamma} (V_{ijkl+}+V_{ijkl-})\right)\;,
\ee  
where 
\be\label{48}
V_{ijkl\pm}:=  E_{ij\pm}\cdot ( E_{ik\pm} \times E_{il\pm})\;.
\ee
Furthermore, up to a sign factor $V_{ijkl\pm}$ does not  depend on the choice of the three disjoint triangles $\{ij\},\{ik\}$ and $\{il\}$ (because of the Gau\ss~constraint and antisymmetry of $C^{ABC}$) and is proportional to the volume squared of the tetrahedron. Note that here the second summand $(V_{ijkl+}-V_{ijkl-})$ vanishes on the geometric subsector, see appendix \ref{appalt}. 

We will not need the other Poisson brackets between the constraints for our line of arguments.   The structure for the Dirac matrix of Poisson brackets $D_{IK}:=\{C_I,C_K\},\, \{C_I\}=\{S_{ij},S_{ij_Ik_I},C_{ijl}\}$ which we determined so far is
\ba\label{49}
\left(\begin{array}{ccc}
0&0&\Fa \\
0&\Fb&\star  \\
-\Fa&\star&\star  
\end{array}\right)\;,
\ea
where $\Fa$ is a diagonal $10\times 10$ matrix giving the Poisson brackets between $S_{ij}$ and $C_{ijl}$. Moreover,  $\Fb$ is the matrix of Poisson brackets between the cross simplicity constraints $S_{ij_Ik_I}$ and has $5$ antisymmetric $2\times 2$ blocks on its diagonal. The inverse is of the following form
\ba\label{49inv}
\left(\begin{array}{ccc}
\star&\star&-\Fa^{-1} \\
\star&\Fb^{-1}&0  \\
\Fa^{-1}&0&0  
\end{array}\right) \; .
\ea  
This already allows us to compute some of the Dirac brackets with respect to the first subset of simplicity constraints defined by
\be\label{50}
\{f,g\}_1=\{f,g\}-\{f,C_K\} (D^{-1})_{KL} \{C_L,g\}  \; .
\ee
A straightforward calculation gives for the Dirac brackets between areas and dihedral angles 
\be\label{51}
\{A_{ij\epsilon},\cos\theta_{i'j'l'\epsilon'}\}_1=\delta_{(ij)(i'j')} \Sigma_{ijk+}\;,
\ee
where $\epsilon, \epsilon = \pm$. Here $\delta_{(ij)(i'j')} = 1$ if the triangles $(ij),(i'j')$ coincide (irrespective of their orientation) and zero otherwise.
From (\ref{51}) we can conclude (see appendix \ref{apppb})
\be\label{51a}
\{a_{ij}\,,\,\, \theta_{i'j'k'}\}_1= \delta_{(ij),(i'j')} \; .
\ee
where $a_{ij}:=2^{-1/2}A_{ij+}^{1/2}$ and $\theta_{ijk}=\theta_{ijk+}$.

The Dirac brackets between two non--opposite 3d dihedral angles in a tetrahedron $\{i\}$ are
\be\label{52}
\{A_{ijk\epsilon},A_{ijl\epsilon'}\}_1=\frac{\gamma}{2} V_{ijkl+}  \;.
\ee
The areas $A_{ij\epsilon}$ still commute with the variables $A_{mnp\epsilon}$.

The results of this section apply not only to the $4$--simplex but to a general 3d triangulation, as the structure of the matrix of Poisson brackets between the first subset of simplicity constraints does not change.

\subsection{Reduction by the gluing constraints}
\label{glue}

We did not implement the full set of simplicity constraints yet, as part of the edge simplicity constraints (\ref{geo5}) are missing. Indeed, the metricity of the triangulation implies gluing constraints, that is constraints involving neighbouring tetrahedra, which we will discuss in this section.

From the last section, we conclude that we can reduce our considerations to the right handed sector, for instance. In the following, we will therefore drop the $\pm$--index.  For the example of a $4$--simplex we are left with $10$ area variables $A_{ij}$, $20$ 3d dihedral angles $\phi_{ij_I,k_I}$ and $10$  4d dihedral angles $\theta_{ijl}$. As is pointed out in \cite{ds}, these variables will not in general be consistent with the geometry of a triangulation. For example, consider two neighbouring tetrahedra $\{i\}$ and $\{j\}$. The four areas and two angles per tetrahedron allow one to compute the length of the edges (for an explicit procedure see \cite{ds}). But there is no guarantee that the lengths of the three shared edges computed with respect to tetrahedra $\{i\}$ and  $\{j\}$ will coincide.     Indeed,   we have to impose the equality of these length variables by further constraints.    Instead of using length variables, one can choose, as in \cite{ds},  the 2d angles $\alpha_{ijkl}$ in the triangle $\{ij\}$ between the edges shared by the triplets of tetrahedra $\{i\}, \{j\}, \{k\}$ and $\{i\},\{j\},\{l\}$, respectively.    Let us denote by $\alpha_{ijkl}$ the angle as computed from the geometrical data associated to the tetrahedron $\{i\}$ and by $\alpha_{jikl}$ the same computed from $\{j\}$.

Then, we have the $30$ gluing constraints
\be\label{53}
\cos\alpha_{ijkl}=\cos\alpha_{jikl} \;,
\ee 
which follow from the fact that the Gau\ss, simplicity, cross simplicity and edge simplicity constraints allow us to conclude that the (dual) bivectors come from the edge vectors of tetrahedra, see also the discussion in \cite{ds}. The $\alpha_{ijkl}$ can be written as a function of the 3d dihedral angles $\phi_{imn}$ 
\be
\cos \alpha_{ijkl} =\frac{N_{ijk}\cdot N_{ijl}}{\sqrt{N_{ijk}\cdot N_{ijk}\,\, N_{ijl}\cdot N_{ijl} }} = \frac{\cos\phi_{ikl}-\cos\phi_{ijk}\cos\phi_{ijl}}{\sin\phi_{ijk}\sin\phi_{ijl}}  \; .
\ee

From these thirty constraints only $10$ are independent (if the Gau\ss~constraints are implemented). This follows from a linearisation of the constraints (\ref{53}) around an equilateral configuration, for example. In fact, it is well known that the internal geometry of a 4--simplex is determined by its $10$ areas or $10$ length variables. (This holds at least locally, as there is a discrete ambiguity in the transformations between areas and length variables, several length assignments can lead to equal area assignments, see \cite{barrettarea}. In the following all considerations apply to local patches of phase space, where this transformation is unique.)

A linearisation of these constraints (\ref{53}) around the equilateral configuration reveals also that it is possible to make a choice of one 2d angle $\alpha_{ijkl}$ per triangle $(ij)$ such that the resulting system is irreducible. 
The Poisson brackets\footnote{or rather the Dirac brackets with respect to the first subset of simplicity constraints (\ref{44}), but we will ignore this terminology and in the following drop the subindex from the bracket.} $\{\,,\}_1$ between the constraints (\ref{53}) can be computed and show that the system is again second class. It is, however, more enlightening to use another but (locally) equivalent set of constraints. Once the constraints (\ref{53}) are imposed, they allow one to consistently compute the length and therefore also the 3d dihedral angles as a function of the areas\footnote{For a non--degenerate triangulation the number of triangles should be at least equal or bigger than the number of edges. Every triangle has three edges, but for a proper 3d piecewise linear manifold every edge should be shared by at least three triangles}. 
Thus, we can replace these constraints by ten constraints of the form
\be\label{54}
G_{iI}=A_{ij_Ik_I}-f_{ij_Ik_I}(A_{mn})\;,
\ee
where the $f_{ij_Ik_I}$ are functions of the squared areas. Unfortunately, we cannot give these functions explicitly. The reason being, that there is not an explicitly known expression for the length as a function of the areas in a 4--simplex, since in order to obtain such an expression, one has to determine the roots of a polynomial equation of order greater than five. We will, however, not need an explicit expression for the $f_{ijk}$ in this discussion.  Another disadvantage is that whereas (\ref{53}) is local in the sense that it involves only variables of two neighbouring tetrahedra we cannot make such a statement for the constraints (\ref{54}). This does not make a big difference for the 3d boundary of a 4--simplex,  but can lead to rather non--local constraints for more complicated triangulations.

The advantage of the constraints (\ref{54}) is the simple structure of their Poisson brackets. The only non--commuting variables appearing in (\ref{54}) are the $A_{ijk}$, so that
\be\label{55}
\{G_{iI}, G_{jJ}\}_1 =  \delta_{ij} \frac{1}{2} V_{il(IJ)k(IJ)m(IJ)} \;,
\ee
where $l(IJ),k(IJ),m(IJ)$ are determined by the choice of indices in the definition (\ref{54}).
Hence the $10\times 10$ Dirac matrix of Poisson brackets of constraints consists of five antisymmetric $2\times 2$ blocks and is invertible.

Since the areas commute with the constraints (\ref{53}) or (\ref{54}), the Dirac brackets  (with respect to the gluing constraints) involving areas do not change from the Poisson brackets $\{\,,\}_1$ defined via the first subset of simplicity constraints (\ref{44}). In other words, the relation (\ref{51}) is ultimately still valid;  the 4d dihedral angles are conjugate to the areas. The areas associated to different triangles still commute with each other. Since we can now express the 3d dihedral angles as functions of the areas (at least locally in phase space), these angles commute with respect to the full Dirac bracket.  We are left with the Dirac brackets between the 4d dihedral angles. From the Jacobi identity, involving two dihedral angles and one area one can conclude that the Dirac bracket between two dihedral angles can be at most a function of the areas. In section \ref{dynamics}, we will conclude by an indirect argument that the 4d dihedral angles are even commuting with respect to the Dirac bracket. This finishes our discussion of the reduced phase space associated to the boundary of a 4--simplex.

\section{The phase space for general triangulations}
\label{general}

Here we will consider general 3d triangulations (of the 3--sphere) and the associated reduced phase spaces. We will assume that these triangulations satisfy the piecewise linear manifold conditions \cite{jan}.
As we will see most of the discussion is very similar to the one for the 4--simplex. In the end, however, we will find that we have additional constraints which involve the areas only and which are, moreover, first class.

We will denote by $N_t$ the number of triangles in the triangulation. Because every triangle is shared by two tetrahedra and every tetrahedron has four triangles, the number of tetrahedra is equal to half the number of triangles $N_t$. The number of phase space variables we start with is $2\times 6\times N_t$. We have $6\times \tfrac{1}{2}\times N_t$ (first class) Gau\ss~constraints, hence the Gau\ss~reduced phase space has dimension $6\times N_t$. We can again choose the gauge--invariant variables from the areas, 3d dihedral angles and 4d dihedral angles as defined in (\ref{41}). More precisely we have two area variables per triangle (for the two $\pm$--sectors) and four independent 3d angle variables per tetrahedron, that is two per triangle. Hence we are again left with two 4d dihedral angles $\theta_{ik,jl\pm}$ per triangle $(ij)$. 

Therefore as a first set of constraints, we have the subset (\ref{44}), which enforce the equality of the quantities as computed in the different chiral sectors. The Dirac matrix of Poisson brackets between the constraints has the same block structure as the one for the 4--simplex and hence is invertible. We will also obtain the same Dirac brackets (\ref{51},\ref{52}) as before.

In a second step we have to impose the gluing constraints (\ref{53}). Again these ensure that the length can be computed consistently from the areas. Hence these constraints allow us again to express also the 3d dihedral angles as a function of the areas, that is to introduce the two constraints $G_{iI},I=1,2$ per tetrahedron. The Dirac matrix of these constraints has again block--diagonal form and is invertible. Note that the original gluing constraints ({53}) are local in the sense that they involve only variables of neighbouring tetrahedra. This is not necessarily the case for the constraints $G_{iI}$.

The $G_{iI}$ constraints do however not exhaust the gluing constraints (\ref{53}). The reason is that for a general 3d triangulation there are more triangles then length variables. 
Hence a consistent geometry also implies constraints $K_{I}(A)$ between the areas. The number of these area constraints is given by the difference of the number of triangles and the number of edges in the 3d triangulation. This suggests that one introduce new configuration variables: namely the constraints $K_{I}(A)$ and length variables $l_{e}(A)$, where the subindex $e$ denotes the edges of the triangulations. Note that in general there are different possibilities to define the length variables as functions of the areas\footnote{These functions are quite complicated if one uses only areas, but it can be more easily defined using the bivectors directly, see appendix \ref{flat}.}-- these choices will differ by terms proportional to the constraints $K_{I}(A)$. One way to find a reduced phase space is to find functions involving also the dihedral angles $\theta$, that are conjugated to the constraints $K_{I}$ and the length variables $l_e$, respectively.

To be more concrete consider the example of the boundary triangulations of two glued 4--simplices. This triangulation has eight tetrahedra and six vertices. To simplify notation we will switch to a vertex based labelling, ie. denote by $v=1,\ldots 6$ the six vertices, $a_{klm}=\sqrt{\frac{1}{2}A_{klm}},\,\,\theta_{klm}$ are the area and dihedral angle variables\footnote{ Here we have to make one choice out of the three possibilities \ref{cons2} for the definition of the dihedral angle. For the final reduced phase space this choice does not matter.} associated to the triangle with vertices $k,l$ and $m$. Assume that we started with two 4-simplices $\sigma_5=(1,2,3,4,5)$ and $\sigma_6=(1,2,3,4,6)$ and glued them together by identifying the tetrahedra $\tau_5,\tau_6$ spanned by the vertices $(1,2,3,4)$ of both 4-simplices.  

After we implemented the Gau\ss~constraint, the first subset of simplicity constraints (\ref{44}) and the constraints $G_{iI}$ we are left with 16 area and 16 4d dihedral angle variables. There are only 14 edges, hence we have two area constraints. These two left--over constraints between the areas can be understood in the following way. From the areas of each 4--simplex we can compute the $3d$ dihedral angles, in particular the ones in $\tau_5$ and $\tau_6$. 
Note that these dihedral angles do not count as variables of our phase space in the first place, as the tetrahedra $\tau_5,\tau_6$ are not part of the 3d boundary of the triangulation. In general we will arrive at different 3d dihedral angles for these two tetrahedra. Indeed the constraints between the areas impose that the 3d dihedral angles in $\tau_5$ and $\tau_6$ have to coincide. This will guarantee that the six edge lengths are independent of the 4-simplex  ($\sigma_5$ or $\sigma_6$) in which they are computed. Since the geometry of a tetrahedron can be parametrised by the four areas and two non--opposite 3d dihedral angles, we end up with two commuting constraints between the areas of the form
\be\label{juli1}
K_I=\phi_I^5(A_{ik5},A_{mno})-\phi_I^6(A_{ik6},A_{mno})  \; , \q\q m,n,o=1,\ldots,4\;,
\ee
which impose that two non--opposite 3d angles (and therefore all 3d angles) in $\tau_5$ and $\tau_6$ coincide.



Define $l_{m5}$ and $l_{m6}$\,, $m=1,\ldots 4$ to be the length between the vertices $m$ and $5$ or $6$ respectively. This length variables can be computed unambiguously from the areas of simplex $\sigma_5$ or $\sigma_6$ respectively. With $l^5_{mn}$ and $l^6_{mn}$,\,$mn=1,\ldots 4$ we will denote the six lengths of the shared tetrahedron as computed from the 4-simplex $\sigma_5$ or 4-simplex $\sigma_6$.

Since the areas $a_{ijk}$ are conjugated to the dihedral angles $\theta_{ijk}$
we will have
\begin{xalignat}{2}\label{juli3}
&\{l_{m5}, p_{l5}\}= \delta_{ml} \;, &&p_{m5}=\sum_{i<j}\frac{\partial a_{ij5}}{\partial l_{m5}} \theta_{ij5} \;,\nn\\
&\{l_{m6}, p_{l6}\}= \delta_{ml} \;, &&p_{m6}=\sum_{i<j}\frac{\partial a_{ij5}}{\partial l_{m6}} \theta_{ij6}\;, \nn\\
&\{l^{5/6}_{mn},p^{5/6}_{m'n'}\}= \delta_{(mn),(m'n')} \;, && p_{mn}^{5/6}= \sum_{i,j,k \neq 6/5,i<j<k} \frac{\partial a_{ijk}}{\partial l_{mn}} \theta_{ijk} \;,
\end{xalignat}
where $m,n,m',n'=1,\ldots,4$.

Now the flow of combinations of the form $p^5_{mn}-p^6_{mn}$ would violate the constraints (\ref{juli1}) as it clearly changes the length variables as computed from the two 4-simplices in different ways. 

The $2\times 4+2\times 6$ momenta $p$ we defined in (\ref{juli3}) are not independent from each other. To find an independent set, we will construct variables conjugated to the $2\times2$ 3d dihedral angles $\phi^5_I$ and $\phi^6_I$.   

To this end consider the change of variables from the six length $\{l^{5/6}_{mn}\}$ of the tetrahedron $\tau_5$ or $\tau_6$ respectively to the four area variables and the two dihedral angles $\{\{a_{ijk}\},\{\phi_I^{5/6}\}\}$ and the associated Jacobi matrix. In this Jacobi matrix, we have the entries 
\be\label{juli5}
{\beta}^I_{mn}:=\frac{\partial l_{mn}}{\partial \phi^I}  \;,
\ee 
where in the partial derivative the areas are kept fixed.
Now we can define 
\be\label{juli6}
\nu_I^{5/6}:=\sum_{m,n\neq 5,6,m<n}\; \sum_{i,j,k\neq 6/5,i<j<k} 
 \frac{\partial a_{ijk}}{\partial l_{mn}} \frac{\partial l_{mn}}{\partial \phi^I} \theta_{ijk} \; .
\ee

Then $\nu^{5/6}_I$ will commute with the four areas of the tetrahedron $\tau_5$ or $\tau_6$ and with the four length $l_{m5}$ and the four length $l_{m6}$.  Moreover $\nu^{5/6}_I$ is conjugated to $\phi_I^{5/6}$. Hence a condition to exclude combinations of momenta whose flow would violate the two area constraints (\ref{juli1}) could be
\be\label{juli6c}
F_I=\nu_I^5-\nu_I^6-c_I\;,
\ee
where $c_I$ is any conveniently chosen constant. 
Conditions such as (\ref{juli6c}), which basically take care of the momenta conjugated to the area constraints (\ref{juli1}), might arise in a proper canonical analysis of the discretised Plebanski action (or area Regge calculus) as secondary constraints, i.e. as the result of taking the Poisson brackets of the area constraints with the Hamiltonian and Diffeomorphism constraints. As long as these conditions form a second class system with the area constraints we do not need their precise form as one can check that the brackets on the reduced phase space involving length variables and the following choice of momenta do not depend on this form. 
This choice of momenta is $p_{l5}, p_{l6}, p_{mn}:=p_{mn}^5+p_{mn}^6$ which are conjugated to $l_{m5},l_{m6},\tfrac{1}{2}(l^5_{mn}+l^6_{mn})$ respectively. These variables commute with the area constraints  and can be taken to parametrise the reduced phase space.

Let us discuss shortly a general triangulation. We will have $(N_t-N_e)$ constraints $K_I(a_t)\,, I=1,\ldots, (N_t-N_e)$ between the area variables $a_t$ associated to the triangles of the triangulation. Here $N_t,N_e$ denote the number of triangles and edges respectively. On the constraint surface $K_I(a)=0$, we can unambiguously define length variables $l_e$ as functions of the areas associated to the edges of the triangulation. Choose some continuation of these functions off the constraint hypersurface. Assuming the length variables to be independent, we can express the areas as functions of the length variables. In this way we can take the derivatives of the constraints $K_I(a_t(l_e))$ with respect to the length variables. Note that these derivatives vanish on the constraint hypersurface. 

 Furthermore we define conjugate momenta $p_e$ associated to the edges as
\be\label{z1}
p_e=\sum_{t @ e} \frac{\partial a_t}{\partial l_e} \theta_t  \;,
\ee
where we sum runs over all triangles hinging on the edge $e$.
Then the Poisson brackets of the momentum $p_e$ with an area constraint $K_I$ gives the derivative of this constraint with respect to $l_e$ and hence it vanishes on the constraint hypersurface. Moreover $p_e$ is conjugated to $l_e$. 

A reduced phase space with respect to the area constraints can be obtained easily since both the length variables and the conjugate momenta commute with the area constraints (at least weakly). Therefore the Poisson bracket on this reduced phase space is given by the Poisson bracket between these quantities on the bigger phase space parametrised by areas and 4d dihedral angles. Also if one introduces another set of constraints (or `gauge conditions') in order to obtain a second class set of constraints $(K_I, F_I)$ the Dirac brackets between length and momenta will coincide with the Poisson brackets on the bigger phase space.

In summary in the final reduced phase space we have length variables and momentum variables associated to the edges of the triangulation. This phase space and momenta (\ref{z1}) can be related to a canonical analysis of Regge calculus \cite{benni}.

One might wonder whether the additional conditions (\ref{juli6}) involving the 4d dihedral angles are not already included in the simplicity, cross simplicity and edge simplicity constraints (\ref{geo4b},\ref{geo5}) we started with. This is actually not the case: one can construct a phase space configuration satisfying all the simplicity constraints with arbitrary 4d dihedral angles, which do not need to satisfy the conditions (\ref{juli6}). 

To this end consider the 3d triangulation embedded in $\Rl^4$. Define edge vectors $e_{ij}=v_j-v_i$ between the vertices $v_i$ and $v_j$. These can be used to construct the bivectors as duals of the wedge products of the edge vectors. That is choose some orientation for every triangle denoted by $t_{ijk}$ and attach the bivector $E_{ijk}$ (with the proper orientation) constructed out of the edge vectors to this oriented triangle.

The holonomies $M_{ijk}$ are chosen to be of the form
\be
M_{ijk}^{AB}=\exp( \lambda_{ijk} C^{ABC}\epsilon^{CD}E_{ijk}^{D}) \; .
\ee
The action of these holonomies on the bivectors can be computed straightforwardly. For instance for $E_{ijk}^{aa'}=\epsilon^{aa'cc'}e_{ij}^ce_{ik}^{c'}$ we have
\ba
C^{aa'bb'cc'}\,(e_{ij}^ce_{ik}^{c'})\,(\epsilon^{aa'cc'}e_{ij}^c b^{c'})=\tfrac{1}{2}\epsilon^{aa'cc'}e_{ij}^c {b'}^{c'}\;,
\ea
with ${b'}^a=2\epsilon^{abcd}e_{ij}^be_{ik}^cb^{d}$. Hence $M_{ijk}$ leaves the edge vectors $e_{ij}$ and $e_{ik}$ invariant. In particular the rotation $M_{ijk}$ leaves the bivector $E_{ijk}$ invariant. Therefore all the constraints (\ref{geo4b},\ref{geo5}) are satisfied. This holds also for the Gau\ss~constraints as these are automatically satisfied for bivectors coming from a tetrahedron. Now the values for the 4d dihedral angles $\theta_{ijk}$ will be proportional to $\lambda_{ijk}$, which can be chosen arbitrarily.


\section{Implementing the Dynamics}\label{dynamics}

So far, we obtained only a kinematical description of simplicial geometries. One way to implement the dynamics is to construct Hamiltonian and diffeomorphism constraints. The crucial question is whether these constraints are first class. Ideally, these constraints should reproduce the space of solutions to Regge calculus in the following sense. Consider the 3d boundary of a 4d triangulation. The 4d triangulation induces canonical data on its boundary, that is, the length of the edges and the 4d dihedral angles. If the 4d triangulation satisfies the Regge equations the associated 3d boundary data should satisfy the constraints. Moreover, a configuration in a gauge orbit of another configuration associated to a 4d solution of the Regge equations, should also be associated to a solution of the Regge equations.

It is not clear if such kind of constraints exist at all. Apart from the fact that we do not have a closed constraint algebra yet for Regge calculus, the main problem is that  a full understanding of diffeomorphisms for general discrete space times is missing. This also includes quantum gravity models involving discrete space or space times, for instance, spin foam models or canonical loop quantum gravity. As is pointed out in \cite{bojo,gft}, the understanding of diffeomorphism symmetry is however crucial for the discussion of anomaly freeness and the appearance of (bubble) divergences in these models. Hence, an understanding of these issues is crucial for the construction and evaluation of quantum gravity models involving discrete space times.
 
Another open problem is the relation between 4d solutions of the Regge equations, the data these solutions induce on the boundary and the dependence on the choice of inner triangulation. In particular, we used so far a `generalised boundary' \cite{oeckl}, namely the 3--sphere, where even for classical and continuous general relativity the question of well--posed initial values is unclear. Furthermore, there is the dependence on the choice of the inner 4d triangulation: consider the 3d boundary of a 4--simplex. A priori, this boundary could be consistent with infinitely many inner triangulations, as one can perform a $1-5$ Pachner move on a 4--simplex, i.e. subdivide this 4--simplex into five 4--simplices. Then, the question is how many solutions to the Regge equations exist if one prescribes for instance the length of the edges in the 3d boundary. One solution, which we will discuss in more detail below is the flat one, corresponding to having only one 4--simplex. (This 4--simplex can be refined into more 4--simplices such that the geometry is still flat.) In fact, in this case, the flat soluiton is the only solution (up to a small number of non-generic solutions which appear to arise as artifacts of the discretisation procedure \cite[benni]). But for more complicated boundaries, we cannot exclude multiple solutions.

To clarify these points, it might be valuable to consider small triangulations, where one could obtain a definite answer. This is work in progress \cite{benni}. Here we will only note that for the simplest non--degenerate case of closed 3d triangulations, namely the boundary of a 4--simplex (and more generally, boundaries of 4d triangulations without \lq inner' triangles) such constraints can be obtained, leading to a flat dynamics. As already noted this might not be the only possibility.
The 4-simplex as a 4d triangulation has no inner triangles, the Regge equations of motion (being linear in the deficit angles attached to the inner triangles) are trivially satisfied. Every 4--simplex can be embedded into 4d flat space time, the dihedral angles for such `flat' 4--simplices can therefore be computed from the length or area variables, see for instance \cite{bdlf3d} for explicit formulae. We will denote by $\Theta_{ij}$ the dihedral angle at the triangle $\{ij\}$ as a function of the areas. Since the dynamics imposes that the 4--simplex is flat, we can take as constraints the 10 functions
\be\label{dyn1}
D_{ij}=\theta_{ij}-\Theta_{ij}  \; .
\ee
(We ignore a sign ambiguity, which encodes the orientation of the 3d boundary of the 4--simplex.) This strategy is very similar to ideas used in \cite{bdlf} for $(2+1)$ dimensional gravity. The commutation relations between the constraints (\ref{dyn1}) are given by
\be\label{dyn2}
\{D_{ij},D_{kl}\}=\{\theta_{ij},\theta_{kl}\}+\gamma\left(\frac{\partial \Theta_{kl}}{\partial a_{ij}} -\frac{\partial \Theta_{ij}}{\partial a_{kl}}\right) \; .
\ee
The second term in brackets vanishes because of the Schl\"afli identity for a 4--simplex. The Schl\"afli identity
\be\label{schlaefli}
\sum_{i,j} a_{ij} \delta \Theta_{ij} =0\;,
\ee
valid for any variations $\delta$ on the space of flat 4--simplices, guarantees that the Regge action for a 4--simplex $\sum_{ij} a_{ij}(\pi-\Theta_{ij})$ is a generating function for the dihedral angles and hence the term in brackets in (\ref{dyn2}) vanishes. 

Also the first term on the right hand side of (\ref{dyn2}) vanishes. As we did not explicitly compute the Dirac bracket between the dihedral angles, we will use an indirect argument. As explained in the Appendix \ref{flat}, one can construct combinations of the flatness constraints whose flow leaves the simplicity constraints invariant on the subspace of flat connections. The geometric interpretation of these constraints is to translate the vertices of the triangulation. Since the flatness constraints are first class this holds also for the vertex translation generators. Moreover, since these combinations of flatness constraints leave the simplicity constraints invariant, there are still first class with respect to the Dirac brackets in the final reduced phase space.

In the following, we argue that the constraints (\ref{dyn1}) are gauge invariant combinations of these flatness constraints and therefore must be first class. Then, we can conclude, that since the Dirac bracket between the dihedral angles $\theta_{ij},\theta_{kl}$ appearing in (\ref{dyn2}) can be at most a function of the areas, 
it has to vanish in order to ensure that the constraints are first class.

Firstly,  the constraints (\ref{dyn1}) implement also the flatness of the (geometric) 4--simplex. Secondly, the action of the constraints (\ref{dyn1}) on the areas is
\be\label{dyn4}
\{a_{kl},D_{ij}\}=\delta_{(kl),(ij)}   \; .
\ee
Hence these constraints generate displacements of the vertices in a way such that only one area at a time changes. Constraints, which change only one length variable at a time can also be constructed
\be\label{dyn5}
\{l_{klm}, \,\, \sum_{i,j} \frac{\partial a_{ij}}{\partial l_{opq}}\theta_{ij}\}=\delta_{(klm)(opq)} \; .
\ee
 
The action of these constraints is to displace one of the vertices of the edge in question in the direction normal to the other three edges adjacent to this vertex. In this way one obtains four displacements per vertex corresponding to the four dimensions of the embedding space--time. This gives $4\times 5$ vertex displacements for the 4--simplex.  Of these 20 displacements, 10 combinations correspond to the 6 global rotations and 4 translations of the 4--simplex. Hence, we have the correct number, 10, of independent constraints.

In summary, for the 3d boundary triangulation of a 4--simplex, we can obtain  first class and even Abelian constraints, which generate the displacement of the vertices. 

This line of arguments can be extended to any 3d triangulation of a 3--sphere which can be taken as the boundary triangulation of a 4d triangulation without inner triangles. Given a 4d triangulation without inner triangles, define  the functions $P'_e(l)$ as derivatives of the Regge action (coinciding in this case with the Hamilton--Jacobi functional) with respect to the length variables. 
\ba
P'_e:=\frac{\partial}{\partial l_e} \sum_t a_t (\pi- \sum_{\sigma @ t} \Theta_t^{\sigma}(l))= \sum_{t @  e} \frac{\partial a_t}{\partial l_e} (\pi - \sum_{\sigma @ t}\Theta_t^{\sigma}(l))\;,
\ea
where in the second sum we sum over the 4--simplices $\sigma$ hinging on the triangle $t$, $\Theta^\sigma_t$ are the dihedral angles in a 4--simplex $\sigma$ expressed as a function of the lengths. The terms with derivatives of the dihedral angles vanish because of the Schl\"afli identity. This suggests that one define momenta conjugate to the length variables by
\be
p'_e:=\sum_{t @  e} \frac{\partial a_t}{\partial l_e} (\pi - \theta_t)\;,
\ee
which only differ by a term involving the areas from the momenta $p_e$ defined in (\ref{z1}) and commute with the area constraints.

The flatness constraints can be brought into a form $D_e=p'_e-P'_e(l_{e'})$ prescribing the momenta $p'_e$ as function of the length variables. The brackets between these constraints are given by
\be
\{D_e,D_{e'}\}=\{p'_e,p'_{e'}\}-\gamma \left(\frac{\partial {P'_{e}}}{\partial l_{e'}} -\frac{\partial {P'_{e'}}}{\partial l_{e}}\right)\;,
\ee
where the second term vanishes because the $P'_e$ come from a generating function. Also, the first term vanishes as the flatness constraints are first class on the BF phase space with which we started. 

Note that such triangulations without inner triangles correspond to a `tree diagram' truncation in group field theory, which is a method to include the sum over all triangulations into a path integral, see \cite{gft}.

Let us count the number of variables and constraints for this type of triangulation. Note that these can be generated by applying $1-4$ Pachner moves starting with the boundary triangulation of a 4--simplex. For every such Pachner move, we add one vertex and four edges to the triangulation. Hence we end up with zero physical degrees of freedom since we expect four (first class) translation--generating constraints per vertex, whereas every edge gives one configuration and one momentum variable.
 
The situation changes if one considers triangulations of the 3--sphere, which cannot be understood as the boundary of 4d triangulations without inner triangles. Such triangulations can be obtained by applying $2-3$ Pachner moves to the 3d triangulation, i.e. by subdividing two neighbouring tetrahedra into three tetrahedra sharing one (new) edge. In the 4d picture, this corresponds to gluing a 4--simplex with two tetrahedra sharing a triangle to the boundary of the triangulation. This triangle turns into an `inner' triangle.

Given such a 3d triangulation with some prescribed length variables, it cannot in general be embedded into flat 4d space. Having performed the $2-3$ Pachner move, in order to keep the induced data 4d flat, the length of the new edge has to be equal to a certain function of the length of the other edges, see also \cite{aristide}. Hence to ensure flatness of the geometry,  one has to include a subset of constraints which are functions of the length only. Since a $2-3$ Pachner move does not lead to new vertices we still have the same number of translation generating constraints as before. With a new edge introduced by a $2-3$ Pachner move, we obtain not only a new length variable (which is fixed by a flatness constraint involving only the length variables) but also a new momentum variable. We can conclude that this momentum variable is also fixed by some flatness constraints, forming a second class system with the constraint which fixes the length of the new edge. In this way we obtain again zero physical degrees of freedom and constraints generating translations of the vertices.\footnote{
Here we differ from `topological gravity' in \cite{wael,zap}: to obtain zero physical degrees of freedom we have to introduce flatness constraints which are functions of the length only and are second class with some other part of the flatness constraints, hence do not generate translations of the vertices. One reason why we might differ from the conclusions in \cite{wael, zap} is that there the Gau\ss~constraints for BF--theory which ensure that the tetrahedra are closed are replaced by Gau\ss~constraints for the triangles, that ensure that the triangles are closed. Yet it remains unclear what kind of gauge transformations these Gau\ss~ constraints based on triangles are generating.
} 

Ultimately one aims of course not for a theory without degrees of freedom, but for one which reproduces solutions of Regge calculus, which in general involve curvature. As we have seen, to impose that the triangulation is 4d flat we have to introduce second class constraints fixing the lengths and momenta of some of the edges. Nevertheless a counting of the degrees of freedom still allows for the possibility to have four translation generating constraints at each vertex. The question is whether one can replace this second class set of constraints and the translation generating constraints by just a set of first class constraints associated to the vertices. This would allow for local degrees of freedom associated to the edges whose length was fixed in the model described before giving flat dynamics.

Another possibility  to impose a dynamics, that we will leave open for further research, is to introduce a discrete time evolution in the form of Pachner moves. Pachner moves on the 3d boundary triangulation can be understood as gluing or removing 4--simplices from the 4d triangulation, hence one can evolve a 3d triangulation and build up in this way a 4d spacetime. These Pachner moves should be implemented as transformations respecting the symplectic structures of the phase spaces the Pachner moves are acting on. As Pachner moves will change the number of tetrahedra, triangles and edges, we have rather to introduce maps between phase spaces based on different 3d triangulations.  

There is an alternative possibility, namely to only allow for combinations of Pachner moves that do not change the connectivity in the 3d triangulation, see \cite{tent}. Triangulation changing Pachner moves and combinations which do not change the triangulations can be realised for (2+1)--dimensional space--times \cite{bdlf}.

\section{Conclusions and Discussion}
\label{conclude}

Let us summarise our line of arguments. We started with the canonical phase space for $SO(4)$ BF--theory on a simplicial (3d) triangulation and considered simplicity, cross simplicity and edge simplicity constraints that enforce the geometricity of the configurations. As a result, on the constraint hypersurface, the $E$--fields are the duals of wedge products of vectors, which we consistently associated to the edges of the triangulation. Moreover, we proved that the constraints are sufficient to ensure geometricity  in Section \ref{proof}. 

Afterwards, we reduced the BF--theory phase space to a phase space describing geometric configurations,  in four steps. As a zeroth step -- to simplify the discussion-- we considered the Gau\ss--reduced phase space. In this way, the rest of the constraints have an immediate geometric interpretation.

In the first step we take care of the constraints (\ref{44}), that guarantee that the areas, 3d dihedral angles and 4d dihedral angles\footnote{Here we actually consider only a subset of the 4d dihedral angles: A priori we can define three angles per triangle and we have to select one of these angles for the parametrisation of the Gau\ss~--reduced phase space. These three angles per triangle coincide on the geometrical subsector. But before the implementation of the gluing constraints \ref{53} the three angles could still be different.} as computed from the right handed sector coincide with the ones computed from the left handed sector.

In the second and third step we dealt with the gluing constraints (\ref{53}). These constraints can be split into two parts. The first set is second class and can be used to solve for the 3d dihedral angles in terms of the area variables. The second set comprises of  constraints between the area variables only and allow the consistent transformation to length variables. This second part is first class. Moreover, we have to get rid of the momenta conjugate to the area constraints. Therefore, we introduce further constraints, turning the first class set onto a second class one, such that the final phase space is parametrised by length variables and conjugated momenta. We conjecture that these further constraints might arise as secondary constraints, that is, from the Poisson brackets of the area constraints with the Hamiltonian and diffeomorphism constraints. 

There are three different forms of Regge calculus, the original one \cite{tregge} based on length variables, area Regge calculus \cite{barrettarea}, which uses areas as fundamental variables but has to impose constraints between the areas, and area--angle Regge calculus \cite{ds} which is based on areas and 3d dihedral angles together with the gluing constraints (\ref{53}). In a canonical analysis, all versions should finally lead to the same reduced phase space based on length variables and conjugated momenta. The kinematical phase spaces of these theories should correspond to the phase space after the first step for area--angle Regge calculus. To obtain the phase space for area Regge calculus one has to solve for the 3d dihedral angles and finally to impose the constraints between the areas and the conjugated conditions involving the 4d dihedral angles to obtain the phase space for length Regge calculus.

Also, the first step of our reduction process leads us to a phase space which corresponds to (the Gau\ss~constraint reduction of) Loop Quantum Gravity restricted to an appropriate fixed triangulation. There the gauge group is $SU(2)$ which corresponds either to the left handed or right handed sector. Performing again a Gau\ss--reduction one would obtain area variables, 3d dihedral angles and 4d dihedral angles. Note however that we did not deal with the full set of geometricity constraints yet. Indeed if one considers a `classical version' of LQG restricted to a fixed triangulation one cannot for instance assign length variables consistently to the edges of this triangulation -- the LQG phase space is truly larger than one describing geometric configuration. One can understand this feature \cite{thiemann} as an enlargement of the space of configurations on which LQG quantisation is based. That is, instead of considering smooth Ashtekar connections (the configuration variable in LQG) one also allows for distributional ones. A recent discussion of this for the quantum theory can be found in \cite{bianchi}. Here we want to point out that there is an explicit set of constraints (\ref{53}) ensuring that length variables can be consistently associated to the edges. Moreover there is a subset, which is second class, hence one cannot impose these constraints exactly via operator equations on the LQG Hilbert space. Of course one can still construct semi--classical states \cite{winkler} that satisfy these constraints to the zeroth order in Planck's constant. 

In the Loop Quantum Gravity phase space the 3d dihedral angles, that is, quantities which one would associate to the intrinsic 3d geometry, do not commute with each other. However, if one implements the gluing constraints via Dirac's procedure one can solve for these 3d angles as functions of the areas and these do commute on the reduced phase space. Hence, the non--commutativity of intrinsic 3d quantities is related to the fact that the phase space is bigger than the one for metric triangulations. 

This enlargement does not appear in the continuum phase space upon which LQG is based. 
More concretely in LQG one usually starts with the cotriads $e^j_a$ as one half of the phase space variables. These are one--forms. It is, however, not possible to obtain a quantum representation for these cotriads -- instead one quantises the two--forms $E^i_{ab}:= \epsilon_{ijk}e^{j}_a e^{k}_b$ (integrated over surfaces). In the continuum, this change of variables can be done without difficulties, i.e. the number of variables does not change. If we consider a fixed triangulation and encode the cotriads into vectors associated to the edges, a change to two--forms integrated over triangles also involves a change of the number of variables and leads therefore to an over--parametrisation of the original phase space based on cotriads. Hence, one can see this enlargement as an result of discretisation. Notice also that a similar issue does not appear in $(2+1)$ dimensional gravity.

Our analysis opens the question of how the dynamics defined, for instance, via the Hamiltonian constraints in loop quantum gravity \cite{hamilt} or the `master constraint' in algebraic quantum gravity \cite{aqg} interacts with the gluing constraints. A natural condition to ask for, is that the Hamiltonian should leave the subsector of geometric configurations invariant at least in an approximate sense.

Simplicity constraints are central for the construction of spin foam models. They ensure that the dynamics imposed is not the one for (topological) BF theory, but the one of general relativity (at least in the gravitational subsector). Also recent work studying the propagation of perturbations in spin foams \cite{carl} show that a correct implementation of the simplicity constraints is crucial in order to obtain the equivalence with general relativity (in this case with Regge calculus). 

On the other hand,  new spin foam models \cite{epr,simet,fk} allowed the matching of boundary data with the LQG Hilbert space (restricted to fixed triangulations) \cite{bdry}. Since in LQG the canonical simplicity constraints are not completely implemented, the question arises of how this can be reconciled with a proper implementation of the simplicity constraints in the path integral, see also the discussion in \cite{alexandrov}. The solution could be in the dynamics, i.e. in the properties of the amplitudes, despite the fact that the simplicity constraints seem not to hold at the kinematical level \cite{lf}.

This brings us to the definition of the (classical) dynamics based on the reduced phase spaces we constructed in our work. 
 To the largest extent we left the dynamics, leading to gravity, open for future research. We proposed two different strategies. One would be to find constraints generating a continuous time evolution in the form of constraints generating translations.  The other would be to find Pachner moves leading to a discrete notion of time evolution. For the former we mainly discussed the possibility to base the translation generating constraints on the vertices of the triangulation. An alternative as proposed in \cite{zap},\footnote{motivated by a counting of degrees of freedom with which we do not completely agree, as explained in section \ref{dynamics},} and used in LQG, is to consider the Hamiltonian as part of the constraints generating translations of the tetrahedra instead. We choose the former possibility as this seems to agree with the counting of degrees of freedom if one wants to impose a flat dynamics. Similar results can be obtained in 3d \cite{bdlf}.

For a very simple subclass of triangulations, in particular for the boundary triangulation for a 4--simplex, one can find a first class set of translation generating constraints leading to a flat dynamics. But already these simple examples could be used to consider mini--superspace reductions along the lines of \cite{rovcos} or study the algebra of Hamiltonian and diffeomorphism constraints along the lines of \cite{bdlf}.

Note that \cite{rovcos} proposes the use of the Loop Quantum Gravity phase space based on small (at least at this stage) triangulations. Since this Loop Quantum Gravity phase space is bigger than the one for metric triangulations\footnote{Although this is not the case for the example considered in \cite{rovcos}, where only two tetrahedra are used to triangulate a 3--sphere. This issue will appear, however, for the boundary of a 4--simplex.}, one can expect additional quantum effects due to these additional degrees of freedom. Here, the simple example of the 4--simplex could be also used to determine whether or not these additional degrees of freedom are propagating.

We did not address the most interesting question here, which is whether it is possible to define a first class set of constraints leading to a non--trivial dynamics with local degrees of freedom. This might not be possible in the end. An alternative in this case is to use Pachner moves, that is to implement a discrete notion of time along the lines of consistent discretisation \cite{gambini}.  

Also ideas from consistent discretisation could be used to start from the discretised action for Plebanski theory and to perform a canonical analysis of this action, as it could clarify some points left open in this work. A continuum analysis has appeared recently in \cite{canpleb} and it would be valuable to know which points have to be altered in a discrete analysis.

\begin{appendix}

\section{Some information on $SO(4)$ and its algebra}\label{def}

The Lie group $SO(4)$ is the group of orthogonal 4x4 matrices of unit determinant, the group of rotations about a fixed point in Euclidean space.  Moreover, its Lie algebra $\so(4)$  is the algebra of antisymmetric 4x4 matrices of zero trace and has elements $J_{aa'}$, where $a, a' = 1,2,3,4$. They satisfy the algebra
\be
[J_{aa'},J_{bb'}] = C^{aa'bb'cc'}J_{cc'},
\ee
where $C^{aa'bb'cc'} = \frac{1}{2}\epsilon^{aa'rs}\epsilon^{bb'st}\epsilon^{dd'tr}\epsilon^{dd'cc'} = \left(\delta^{ab}\epsilon^{a'b'cc'} + \delta^{a'b'}\epsilon^{abcc'} - \delta^{a'b}\epsilon^{ab'cc'} - \delta^{ab'}\epsilon^{a'bcc'}\right)$ and $\epsilon^{abcd}$ is the totally antisymmetric Levi-Civita tensor.

We utilise in the paper some labour saving notation.  We denote the antisymmetric combination of Lorentz indices by capital letters from the beginning of the Latin alphabet 
\be
T^A := T^{[aa']}\;,\quad\quad\text{where $T$ is an arbitrary tensor.}
\ee
In particular, we define
\be
\delta^{AB} := 2\delta_{[a}^{[b}\delta_{a']}^{b']}\quad\text{and}\quad \epsilon^{AB} := \epsilon^{aa'bb'}.
\ee
On top of this, we also invoke an idiosyncratic summation convention for these indices, namely
\be
T_1^AT_2^A = \frac{1}{2}T_1^{aa'}T_2^{aa'}\;.
\ee
The upshot of these definitions is that
\be
\epsilon^{AB}\epsilon^{BC} = \delta^{AC} \q \text{and}\q  T^A=\delta^{AA'}T^{A'}\; .
\ee
Also, the Hodge dual on the algebra is defined as
\be
{}^*E^A = \epsilon^{AB}E^B \; .
\ee
Furthermore, the following identities hold
\ba
\epsilon^{AA'}C^{A'BC} &=& \epsilon^{BB'}C^{AB'C} = \epsilon^{CC'}C^{ABC'}\;,\\
C^{ABC} &=& M^{AA'}M^{BB'}M^{C'C}C^{A'B'C'}\;,\\
\epsilon^{AA'}M^{A'B}&=&\epsilon^{BB'}M^{AB'}\;,\\
(M^{-1})^{DA}&=&M^{AD}\;.
\ea
for $M\in SO(4)$, and we define a anti-symmetric product of bi-vectors as:
\be
(T_1 \times T_2)^A := C^{ABD}\epsilon^{DC} T_1^B T_2^C\;.
\ee

The Lie algbera splits into selfdual and anti-selfdual parts.  Explicitly, we invoke projectors
\be\label{can6}
P_{\pm}^{AA'}:=\frac{1}{2}(\delta^{AA'}\pm \epsilon^{AA'})\;.
\ee 
It is straightforward to check that these projectors are orthonormal, that is $P_s^{AB}P_{s'}^{BC}=\delta_{ss'}P_s^{AC}$, furthermore $P_+^{AB}+P_-^{AB}=\delta^{AB}$. 
We may then proceed to generate all manner of projected quantities
\ba
E_\pm^A &=& P_\pm^{AA'} E^{A'}\;,\quad\quad\quad\quad M_\pm^A = P_\pm^{AA'}M^{A'}\;,\\
C_\pm^{ABC} &=& P_{\pm}^{AA'}C^{A'BC} = P_{\pm}^{BB'}C^{AB'C} = P_{\pm}^{CC'}C^{ABC'} = P_{\pm}^{AA'}P_{\pm}^{BB'}P_{\pm}^{CC'}C^{A'B'C'}\;.
\ea
The following identity for the $SO(4)$ structure constants 
\be\label{cons1}
C^{ABC}_\pm C^{A'B'C}_\pm=2^3 (P^{AA'}_\pm P^{BB'}_\pm - P^{AB'}_\pm P^{A'B}_\pm ).
\ee
 is in close analogy to the relation $\epsilon^{abc}\epsilon^{a'b'c}=(\delta^{aa'}\delta^{bb'}-\delta^{ab'}\delta^{a'b})$ for the $SO(3)$ structure constants.


\section{The discrete BF-theory}\label{discrete}
Our continuum action is
\be
{\cal S}[E,A]  =  \int (E+\frac{1}{\gamma} {}^* E)\wedge F \; .
\ee
We rewrite this in the Hamiltonian formalism
\be
{\cal S}[\Pi, A] = \int dt\int_\Sigma d^3x \;\epsilon_{ijk}\left( \Pi_{ij}^A A_{k}^A + \Pi_{0i}^A F_{jk}^A  + A_0^A D_{i}\Pi_{jk}^A\right) \;,
\ee
where $\Pi := E+ \frac{1}{\gamma} {}^*E$, the indices $i,j,k$ denote components of the forms on the spatial hypersurface $\Sigma$, and $\epsilon_{ijk} := \epsilon_{0ijk}$.  The components $\epsilon_{ijk}\Pi_{jk}$ are the canonical momenta to the $A_i$, while the components $\Pi_{0i}$ are Lagrange multipliers enforcing the flatness constraint $\epsilon_{ijk}F_{jk} = 0$. Also, $A_0$ is a further Lagrange multiplier imposing the Gau\ss~constraint $\epsilon_{ijk}D_i\Pi_{jk} := \epsilon_{ijk}(\partial_i\Pi_{jk} + [A_i,\Pi_{jk}])= 0$, where $[\,,]$ denotes the commutator on the $\so(4)$ Lie algebra.  As expected in a parametrisation invariant theory, the Hamiltonian is just the sum of constraints.  The Poisson brackets between the canonical momenta are
\be
\{A_i^A(\vec{x}), \Pi_{jk}^B(\vec{y})\} = \epsilon_{ijk}\delta^{AB}\delta^{3}(\vec{x}-\vec{y}) \;.
\ee
The discussion of the canonical phase space for discrete BF--theory follows closely \cite{wael}. In addition, we have to deal with the Barbero--Immirzi parameter. We discretise our 3d spatial hypersurface $\Sigma$ using a simplicial lattice $\Delta$.  Such a simplicial complex contains tetrahedra $T$, triangles $t$, edges $e$ and vertices $v$. We shall denote the tetrahedra by the Latin indices $i,j$ etc.  Using this, we can pick out the triangle $ t = \{ij\}$, with respect to which the tetrahedra $i$ and $j$ are adjacent.   On top of this, we may pinpoint an edge $e = \{ijk\}$ which is common to the triangles $\{ij\}$ and $\{ik\}$.   Furthermore, we define the topological dual $\Delta^*$ of the simplicial complex to be that structure which associates a $(3-n)$--dimensional object to each $n$--dimensional subsimplex of $\Delta$.  Thus, we denote the vertices of $\Delta^*$ by $i,j,..$ and the edges of $\Delta^*$ by pairs $\{ij\}$.   The spatial connection and its canonical momentum possess natural discrete analogues in this setting
\ba
	A &\rightarrow& A_{ij} = ln\; M_{ij} := ln \; \int_{ij} A \quad\in \so(4)\;,\\
	\Pi &\rightarrow& \Pi_{ij} := \int_{ij} \Pi \quad\in\so(4)\;,
\ea
where $M_{ij}$ is the parallel transport matrix from tetrahedron $i$ to tetrahedron $j$.  We may then re-express 
\be
	\dot{A} \rightarrow  \dot{A}^A_{ij} = \frac{1}{2}C^{ABC} [M_{ij}^{-1}]^{AD} \dot{M}_{ij}^{DC}.
\ee
The Lagrange multipliers do not possess natural discrete analogues since we are not discretising the full space time manifold.    We write the discrete action as
\be
	{\cal S}[\Pi_{ij}, M_{ij}] = \int dt\left[ \sum_{i,j} C^{ABC} \Pi_{ij}^A [M_{ij}^{-1}]^{BD}\dot{M}_{ij}^{DC} + \sum_{e}N_e^A G_e^A + \sum_{i}\tilde{N_i}^A \tilde{G}_i^A\right](t)\;,
\ee
where the flatness and Gauss constraints are replaced by their respective discrete counterparts
\ba 
	G_{e}^A & = & C^{ABC} \left(M_{ij}M_{jk}\dots M_{ni}\right)^{BC},\label{flatnessconstraint}\\
	\tilde{G}_i^A & = & \sum_{k}\Pi_{ik}^A.
\ea
Thus our fundamental phase space variables are $(M_{ij}^{AB}, \Pi_{ij}^C)$ associated to the triangles $(ij)$.  Alternatively, we may take $(M_{ij}^{AB}, E_{ij}^C )$ where $E_{ij}^A = \frac{\gamma^2}{\gamma^2-1}(\Pi - \frac{1}{\gamma}{}^*\Pi)_{ij}^A$.
They satisfy the following commutation relations
\ba
	\{E^A_{ij},M_{kl}^{BC}\}&=&\frac{\gamma^2}{\gamma^2-1}\,(\delta^{AA'}-\frac{1}{\gamma}\epsilon^{AA'})\left(\delta_{ik}\delta_{jl}\,C^{A'BD}M_{kl}^{DC} + \delta_{il}\delta_{jk} \,C^{A'CD}M_{kl}^{BD}\right),\quad\quad \\
	\{E^A_{ij},E^B_{kl}\}&=&\delta_{ik}\delta_{jl} \frac{\gamma^2}{\gamma^2-1}C^{ABC}(\delta^{CC'}-\frac{1}{\gamma} \epsilon^{CC'})E_{kl}^{C'}\;.
\ea
The second equation may be somewhat puzzling since the momenta commute in the continuum theory, but this structure is natural in a theory where the momentum space is the cotangent space to a group manifold, in our case $SO(4)$.  In any case, such a commutation relation is necessary to satisfy the Jacobi identity
\be
\{\{E^A,E^B\},M^{CD}\} +\{\{E^B,M^{CD}\},E^A\}+ \{\{ M^{CD},E^A \}, E^B\} = 0\;,
\ee
where we suppressed the subindices as we consider here only variables defined on one and the same triangle.  Furthermore, they are subject to the conditions
\ba
	\label{tri} 		E_{ji}^A &=& -M_{ji}^{BA}E_{ij}^B\;,\\ 
				M_{ij}^{AC}M_{ji}^{CB} &=& \delta^{AB}\;,\\
				M_{ij}^{AC}M_{ij}^{BC} &=& \delta^{AB}\;,
\ea 
which encapsulate the properties that the $E$ field when viewed from one tetrahedron is rotated when viewed from the other, that the parallel transport matrix $M$ is mapped to its inverse under a change of orientation, and that $M$ is orthogonal.

\section{The 4d dihedral angles}
\label{form}

We defined the (moduli of the) 4d dihedral angles $\theta_{\pm}$ in (\ref{cons2}) by
\be\label{sun1}
\cos\theta_{ik,jl\pm}=\frac{N_{ijk\pm} \cdot (M_{ij}N_{jil})_\pm}{\sqrt{N_{ijk\pm}\cdot N_{ijk\pm}\,\, N_{jil\pm}\cdot N_{jil\pm}}} \; .
\ee
We can obtain  an expression for $\sin\,\theta_\pm$ by using $(\sin\,\theta_\pm)^2 = 1- (\cos\,\theta_\pm)^2$.   By virtue of the contraction of two structure constants (\ref{cons1}) we can write
\ba\label{cons3}
\frac{
(N_{ijk\pm} \times (M_{ij}N_{jil})_\pm)\,\cdot\,(N_{ijk\pm} \times (M_{ij}N_{jil})_\pm)
}
{2^3\, N_{ijk\pm}\cdot N_{ijk\pm}\,\, N_{jil\pm}\cdot N_{jil\pm} }
&=&
1- \frac{   (N_{ijk\pm}\cdot (M_{ij} N_{jil})_\pm )^2     } { N_{ijk\pm}\cdot N_{ijk\pm}\,\, N_{jil\pm}\cdot N_{jil\pm} }  \q,\q\q 
\ea
where we used the notation $(T_1\times T_2)^A=C^{ABD}\epsilon^{DC}T^B_1 T^C_2$. 
The left hand side of (\ref{cons3}) is therefore equal to $(\sin\theta_\pm)^2$. The expression on the left hand side simplifies because of
\ba\label{cons4}
	(N_{ijk\pm} \times (M_{ij}N_{jil})_\pm)&=& 2^3 \, \left(E_{ij\pm}\cdot (M_{ij} N_{jil})_\pm\,\, E_{ik\pm}- E_{ik\pm} \cdot (M_{ij}N_{jil})_\pm \,\, E_{ij\pm}\right)\;,\nn\\
	&=& -2^3\,\, E_{ik\pm} \cdot (M_{ij}N_{jil})_\pm \,\, E_{ij\pm}\;  ,
\ea
since the first term in the first line of (\ref{cons4}) vanishes due to the contraction of $E_{ij\pm}$ with $N_{jil\pm}$. Hence we have
\be\label{cons5a}
\sin^2\theta_{ik,jl\pm}=\frac{2^3\, E_{ij\pm} \cdot E_{ij\pm} \,\, ( E_{ik\pm} \cdot (M_{ij}N_{jil})_\pm )^2}
{N_{ijk\pm}\cdot N_{ijk\pm}\,\, N_{jil\pm}\cdot N_{jil\pm}   }  \; .
\ee
An expression that will appear later on in the Poisson bracket calculations is
\ba\label{cons5}
 E_{ij\pm}  \cdot (N_{ijk\pm} \times (M_{ij}N_{jil})_\pm) 
&=&
C^{ABC}_\pm C^{BDE}_\pm E_{ij\pm}^A E_{ij\pm}^D E_{ik\pm}^E (M_{ij}N_{jil})_\pm^C\;, \nn\\
&=&-2^3 E_{ij\pm}\cdot E_{ij\pm}\,\,\, E_{ik\pm} \cdot (M_{ij}N_{jil})_\pm \;,\nn\\&=&
-2^{3/2} \sin\theta_{ik,jl\pm} \,\,\,|E_{ij\pm}|\,\, |N_{ijk\pm}|\,\,|N_{jil\pm}|  \; , 
\ea
where $|T|=\sqrt{T\cdot T}$. Note that in the last line of (\ref{cons5}) we also defined the signs for $\theta_{ij,kl\pm}$.


\section{Alternative geometricity constraints}\label{appalt}

The simplicity constraints can be partially replaced by alternative constraints, which also distinguish between the topological and gravitational sector. Consider the totally antisymmetric expressions
\ba
V(E,E',E''):=C^{ABD}\epsilon^{DC}E^A{E'}^B{E''}^C  \q \text{and}\q\q W(E,E',E''):=C^{ABC}E^A{E'}^B{E''}^C  \; .
\ea
in the three bivectors $E,E',E''$. Under the duality map $E\mapsto {}^*E$ the expression $V$ is mapped to $W$ and vice versa. It is straightforward to check that $V$ vanishes for a triple of bivectors of the form 
\ba\label{sun4c}
E={}^{*}(e\wedge e') \; ,\q
E'={}^*(e\wedge e'') \; , \q
E''= {}^*(e\wedge e''') \;,
\ea
whereas $W$ vanishes for
\ba
E={}^{*}(e\wedge e')\;,\q
E'={}^{*}(e'\wedge e'') \; ,\q  
E''={}^{*}(e''\wedge e)  \; .
\ea

Hence we have for instance $V(E_{ij+},E_{ik+},E_{il+})=V(E_{ij-},E_{ik-},E_{il-})$ on the gravitational sector. Note that due to the Gau\ss~constraint, $V(E_{ij},E_{ik},E_{il})$ does not depend (modulo a sign) on the choice of the three triangles of the tetrahedron. Indeed it is proportional to the square of the volume of the tetrahedron.

On the gravitational sector three bivectors sharing an edge (and transported to the same coordinate system) are of the form (\ref{sun4c}). Hence we have the constraints
\be\label{sun5a}
   E_{ik}^A\, M_{ij}^{AB}  C^{BCE}\epsilon^{ED}  E^C_{ji} E_{jl}^D=E_{ik}\cdot (M_{ij}N_{jil})=0\;.
\ee
if the three dual bivectors involved share an edge. Indeed these constraints are used in \cite{wael} instead of the edge simplicity constraints. 
Note that 
\be\label{sun6}
0=E_{ik+}\cdot (M_{ij}N_{jil})_+ +E_{ik-}\cdot (M_{ij}N_{jil})_-\;,
\ee
so that together with (\ref{cons5}) we have $\sin\theta_{ik,il+}=-\sin\theta_{ik,il-}$ on the gravitational sector (and if the three triangles involved in the definition of $\theta_{ij,kl}$ share an edge).

\section{Poisson brackets}\label{apppb}

Here we will compute in more detail the Poisson brackets (\ref{45},\ref{47}). To obtain the Poisson brackets between the area squared variables $A_{ij}$ and the cosine of the 4d dihedral angles we compute
\ba\label{sun2}
&&~ \!\!\!\!\!\! \!\!\!\!\!\! 
 \!\!\!\!\!\! \!\!\!\!\!\! 
 \!\!\!\!\!\! \!\!\!\!\!\! 
 \!\!\!\!\!\! \!\!\!\!\!\! 
\left\{E_{ij\pm}\cdot E_{ij\pm}\,,\,\, \frac{N_{ijk\pm} \cdot (M_{ij}N_{jil})_\pm}{\sqrt{N_{ijk\pm}\cdot N_{ijk\pm}\,\, N_{jil\pm}\cdot N_{jil\pm}}}\right\} \nn\\
&=&
 \frac{2 E_{ij\pm}\cdot }
{\sqrt{N_{ijk\pm}\cdot N_{ijk\pm}\,\, N_{jil\pm}\cdot N_{jil\pm}}}
 \,\, \{ E_{ij\pm}\,,\,\,  N_{ijk\pm} \cdot (M_{ij}N_{jil})_\pm  \}\;,  \nn\\
&=&\pm\frac{2\gamma}{\gamma \pm 1}  \frac{ E_{ij\pm}\cdot  (N_{ijk\pm} \times (M_{ij}N_{jil})_{\pm}))} 
{\sqrt{N_{ijk\pm}\cdot N_{ijk\pm}\,\, N_{jil\pm}\cdot N_{jil\pm}}} \;, \nn\\
&=:&\pm\frac{2\gamma}{\gamma\pm 1} \Sigma_{ik,jl\pm}\; .
\ea
Here, we used that $E_{ij}\cdot E_{ij}$ commutes with both $N_{ijk}$ and $N_{jil}$. Thus, the area squared variable $A_{ij}$ commutes with all 4d dihedral angles with the exception of the ones associated to the triangle $\{ij\}$. As a result,  the Poisson brackets between the constraints $S_{ij}$ and $C_{ik,jl}:=\cos\theta_{ij,kl+}-\cos\theta_{ik,jl-}$ are
\be
\{S_{ij}, C_{ik,jl}\}\simeq -4 \frac{\gamma^2}{\gamma^2-1} \Sigma_{ik,jl+}\;,
\ee
where the equality sign $\simeq$ indicates that we evaluate the expression on the gravitational sector, where we have $\Sigma_{ik,jl+}=-\Sigma_{ik,jl-}$ because of equations (\ref{cons5},\ref{sun6}).   The Dirac bracket $\{A_{ij\epsilon},\cos\theta_{ik,jl\epsilon'}\}_1$ as defined in the main text can be computed straightforwardly to be
\ba
\{A_{ij\epsilon}, \cos\theta_{ik,jl\epsilon'}\}_1
&\simeq& \Sigma_{ik,jl+} \;, \nn\\ 
&\simeq&-2^{3/2} \,\sin\theta_{ij,kl+} \sqrt{A_{ij+}}  \; .
\ea
From this we can conclude that the Dirac bracket between $a_{ij}:=2^{-1/2}\sqrt{A_{ij+}}$ and $\theta_{ik,jl}:=\theta_{ik,jl+}$ is given by
\be
\{a_{ij}\,,\,\, \theta_{ik,jl}\}_1=1 \; .
\ee
The Poisson brackets between two non--opposite angle variables $A_{ijk\pm},A_{ijl\pm}$ can be computed to
\ba
\{E_{ij\pm}\cdot E_{ik\pm}\,,\,\, E_{ij\pm}\cdot E_{il\pm}\}
&=& \pm\frac{\gamma}{\gamma\pm 1} E_{ij\pm}\cdot( E_{ik\pm}\times E_{il\pm})\;,\nn\\
&=:& \pm\frac{\gamma}{\gamma\pm 1} V(E_{ij\pm},E_{ik\pm},E_{il\pm})  \; .
\ea
Now on the gravitational sector we have $V(E_{ij+},E_{ik+},E_{il+})=V(E_{ij-},E_{ik-},E_{il-})=:V_{ijkl}  $ and therefore
\ba
\{S_{ijk},S_{ijl}\}\simeq -\frac{2\gamma}{\gamma^2-1} V_{ijkl}  \; .
\ea
This gives the following Dirac brackets
\ba
\{A_{ijk\epsilon},A_{ijl\epsilon'}\}_1\simeq -\frac{\gamma}{2}V_{ijkl}\;.
\ea
between the variables $A_{ijk\epsilon}$ and $A_{ijl}$ based on non--opposite edges in the tetrahedron $\{i\}$.

\section{Flat Dynamics in the BF theory phase space}\label{flat}

Adapting \cite{wael} to our purposes we want to construct combinations of the flatness constraints (\ref{flatnessconstraint}) that leave the simplicity constraints and therefore the sector of geometric configurations invariant. 

We will use the slightly altered flatness constraints  
\ba
F^A_{ei}:=2^{-4}C^{ABC}\epsilon^{CD}(\delta^{DE}+\frac{1}{\gamma}\epsilon^{DE})(M_{ij}M_{jk}\cdots M_{ni})^{BE}\;,
\ea
where $e$ is the edge around which the holonomy is taken and the subindex $i$ denotes the tetrahedron at which the holonomy starts and ends. For a bivector $E_{lm}$ associated to a triangle $\{lm\}$ (oriented in the direction of the loop) sharing the edge $e$ the action of the constraint is given by
\ba
\{E^A_{lm},\Lambda^CF^C_{ei}\} \simeq  \, \Fm_{li}^{AC'} \Lambda^{C'}\;,
\ea
where the equality sign $\simeq$ indicates that we evaluate the result on the subspace of flat connections and $\Fm_{li}$ is (some) parallel transport from the tetrahedron $\{i\}$ to the tetrahedron $\{l\}$.

Now one can define combination of flatness constraints associated to the edges hinging at a vertex $v$, such that the corresponding action translates this vertex at least on the gravitational and flat subsector. 

To this end one has to find a function that gives (on the gravitational subsector) the vector $e^a_e$ associated to an edge $e$ as a function of the bivectors from one of tetrahedrons sharing this edge. (Unfortunately the formulae given in \cite{wael} are incorrect. The antisymmetrisations appearing in the volume form proposed there lead to a vanishing expression.) We can construct this edge vector by calculating its direction and length.  The bivectors in question are 
\be
E_{ij} = {}^*(e_1\wedge e_2)\;,  \quad\quad E_{ik} = {}^*(e_2\wedge e_3)\;,\quad\quad E_{il} = {}^*(e_3\wedge e_1)\;,
\ee
 for three faces of the tetrahedron $\{i\}$.  The length of the vector $e_1$ is 
 \be
 |e_1| = \sqrt{\frac{N_{ijl}\cdot N_{ijl}}{2|E_{ij}\cdot (E_{ik}\times E_{il})|}}\; .
 \ee
 since the numerator is  $(length\times volume)^2$ while the denominator is $(volume)^2$.
 Uncovering its direction is a somewhat more arduous calculation. We begin by defining
 \be
 \ca^{abc} = N^{aa'}_{ijk}N^{bb'}_{ikl}N^{cc'}_{ijl}x^{d'}\epsilon^{a'b'c'd'} = \left(e_1^{a'}e_2^{b'}e_3^{c'}x^{d'}\epsilon^{a'b'c'd'}\right) n^a_\tau n^b_\tau n^c_\tau +\left\{\text{\small terms proportional to $e^a n^b_\tau n^c_\tau$}\right\},
 \ee
 where we have chosen an auxiliary vector vector $x^{a}$, transverse to the tetrahedron.  We then use this to produce
 \be
 \cb^{abc} = \left(N^{[a|a'|}_{ijk}N^{b|b'|}_{ikl}N^{c]c'}_{ijl}\right) \ca^{a'b'c'}  = \left(e_1^{a'}e_2^{b'}e_3^{c'}x^{d'}\epsilon^{a'b'c'd'}\right) e^{[a}_1\,e^b_2\,e^{c]}_3 +\left\{\text{\small terms proportional to $e^{[a} e^b n^{c]}_\tau$}\right\}
 \ee 
 Finally, the vector $e^a_1$ can be picked out of $\cb^{abc}$, up to a multiplicative factor, by contracting it appropriately with the dual bivectors ${}^*E_{ij}$ and ${}^*E_{il}$.  This results in 
\be
 \frac{e_1^a}{|e_1|} = \frac{B^{bcd}\,{}^*E^{ab}_{ij}\,{}^*E^{cd}_{il}}{\sqrt{(B^{bcd}\,{}^*E^{ab}_{ij}\,{}^*E^{cd}_{il})(B^{b'c'd'}\,{}^*E^{ab'}_{ij}\,{}^*E^{c'd'}_{il})}}\;,
\ee 
where the denominator on the right hand side normalises the resulting vector.  Note that we have its length and direction, we know everything we need to reconstruct the vector $e_1$.

With these function we define the constraints associated to a vertex
\ba\label{mon1}
H_a(v,i)=\sum_{e @ v} \epsilon_{abcd} e_{ei}^b F_{ei}^{cd}\;,
\ea
where  $e_{ei}^a$ are the edge vectors expressed in the coordinates of the tetrahedron $\{i\}$  and we sum over all edges adjacent to $v$ and oriented towards $v$. The orientation of the loops on which the holonomies are based on and the associated edges are positive. Note that the Poisson algebra of these constraints closes on the hypersurface defined by the vanishing of the flatness constraints $G_{e'}=0$ for all edges $e'$.

The constraint (\ref{mon1})  has the following action on a bivector ${}^*E_{kl}^{aa'}=e_{ek}^{[a}e_{e'k}^{a']}$, with $e,e'$ edges adjacent and pointing to $v$:
\ba
\{ E_{kl}^{gg'}, \lambda^a H_a(v,i)\} &\simeq& \Fm_{ik}^{bb'aa'} \lambda^{[a}_{~} e^{a']}_{\;\;ei}\,\, \epsilon_{bb'}^{\;\;\;\;gg'} - \Fm_{ik}^{bb'aa'} \lambda^{[a}_{~} e^{a']}_{\;\;e'i}\,\, \epsilon_{bb'}^{\;\;\;\;gg'}\;.
\ea

This is exactly the change the bivector $E_{kl}$ undergoes under a translation of the vertex by a vector $\lambda$ (in the coordinate system $\{i\}$). Hence the vertex translation constraints (\ref{mon1}) leave the subsector of geometric configurations invariant. Note however that the entire construction works only on the subspace of flat holonomies.

\section{The LQG phase space}

The phase space corresponding to loop quantum gravity restricted to a fixed 3d triangulation coincides with the one for $SU(2)$--BF theory. That is, we can apply the same notations and conventions as for the $SO(4)$--BF theory, we have only to change the $SO(4)$--labels $A,B,\ldots$ to $SU(2)$--labels $a,b,\ldots$ taking values $1,2,3$. The Poisson brackets between the basic fields are \cite{tt}
\ba
\{E_{ij}^a,M_{ij}^{bc}\}&=&\gamma \epsilon^{abd}M^{dc} \;, \nn\\
\{E_{ij}^a, E_{ij}^b\}&=&\gamma\epsilon^{abc} E_{ij}^c\;,
\ea
and so on, with $\epsilon^{abc}$ the totally antisymmetric Levi--Civita tensor.

The Gau\ss~constraints
\be
\sum_{k} E_{ik}^a=0\;,
\ee
generates $SU(2)$ gauge transformations. Gauge invariant quantities can be defined analogously to the $SO(4)$ case
\ba
A_{ij}&:=&E^a_{ij}E^a_{ij}\;, \nn\\
A_{ijk}&:=&E^a_{ij}E^a_{ijk}\;,\nn\\
\cos\theta_{ik,jl}&:= &\frac{N^a_{ijk}M^{ab}_{ij}N^b_{jil}}{\sqrt{N_{ijk}^d N^d_{ijk}\,\, N^e_{ijl} N^e_{ijl} }}\;,
\ea
where $N_{ijk}^a:=\epsilon^{abc}E_{ij}^bE^c_{ik}$. A counting of degrees of freedom in the gauge invariant phase space gives the same result as the one for say the right handed sector of $BF$--theory, that is we have one area variable $A_{ij}$ per triangle, two 3d dihedral angle variables $A_{ijk}$ per tetrahedron and we have to choose one $\theta_{ik,jl}$ per triangle. Again this does not necessarily mean that the three angles associated to one triangle coincide, just that there are relations between these 4d angles which may involve the 3d angles.

Also the symplectic structure will be analogous to (\ref{51a},\ref{52}). That is the areas are conjugated to the dihedral angles $\theta$, which encode the extrinsic curvature. Note that here $\{a_{ij},\theta_{ijk}\}_{\text{LQG}}=\gamma$ as compared to (\ref{51a}), which gives $\{a_{ij},\theta_{ijk}\}=1$. The reason is that in LQG one uses the Ashtekar connection in which the extrinsic curvature appears multiplicated with the Immirzi parameter $\gamma$.  One therefore needs to rescale the angles $\theta$ by $1/\gamma$ to obtain the true values for the extrinsic curvature.

\end{appendix}

\vspace{0.5cm}

\subsection*{Acknowledgements}

We  thank Benjamin Bahr, Florian Conrady, Laurent Freidel, Lee Smolin, Thomas Thiemann and especially Sergei Alexandrov as well as Simone Speziale for discussions. Research at Perimeter Institute is supported by the Government of Canada through Industry Canada and by the Province of Ontario through the Ministry of Research and Innovation.


\begin{thebibliography}{99}
\parskip -2pt



\bibitem{ds} 
  B.~Dittrich and S.~Speziale, 
  ``Area-angle variables for general relativity,'' to appear in New Journal of Physics
  [arXiv:0802.0864 gr-qc].


\bibitem{reviews}
  A.~Ashtekar and J.~Lewandowski,
  ``Background independent quantum gravity: A status report,''
  Class.\ Quant.\ Grav.\  {\bf 21} (2004) R53
  [arXiv:gr-qc/0404018].\\
   C.~Rovelli, ``Quantum Gravity'', (Cambridge University Press, Cambridge 2004)\\
T.~Thiemann,``Modern canonical quantum general relativity'', (Cambridge University Press, Cambridge 2007) 


\bibitem{regge}
  T.~Regge,
  ``General relativity without coordinates,''
  Nuovo Cim.\  {\bf 19}, 558 (1961).


\bibitem{williams}
  T.~Regge and R.~M.~Williams,
  ``Discrete structures in gravity,''
  J.\ Math.\ Phys.\  {\bf 41} (2000) 3964
  [arXiv:gr-qc/0012035].

\bibitem{cdt}
  J.~Ambjorn, J.~Jurkiewicz and R.~Loll,
  ``Quantum gravity, or the art of building spacetime,''
  arXiv:hep-th/0604212.


\bibitem{hamilt}
  T.~Thiemann,
  ``Anomaly-free formulation of non-perturbative, four-dimensional  Lorentzian
  quantum gravity,''
  Phys.\ Lett.\  B {\bf 380} (1996) 257
  [arXiv:gr-qc/9606088].\\
  T.~Thiemann,
  ``QSD V: Quantum gravity as the natural regulator of matter quantum field
  theories,''
  Class.\ Quant.\ Grav.\  {\bf 15} (1998) 1281
  [arXiv:gr-qc/9705019].


\bibitem{finiteness}
  L.~Crane, A.~Perez and C.~Rovelli,
  ``A finiteness proof for the Lorentzian state sum spinfoam model for  quantum
  general relativity,''
  arXiv:gr-qc/0104057.


\bibitem{bojo}
  M.~Bojowald and A.~Perez,
  ``Spin foam quantization and anomalies,''
  arXiv:gr-qc/0303026.









\bibitem{friedman}
  J.~L.~Friedman and I.~Jack,
  ``(3+1) Regge Calculus With Conserved Momentum And Hamiltonian Constraints,''
  J.\ Math.\ Phys.\  {\bf 27} (1986) 2973.


\bibitem{piran}
  T.~Piran and R.~M.~Williams,
  ``A (3+1) Formulation Of Regge Calculus,''
  Phys.\ Rev.\  D {\bf 33} (1986) 1622.


\bibitem{loll}
  R.~Loll,
  ``On the diffeomorphism-commutators of lattice quantum gravity,''
  Class.\ Quant.\ Grav.\  {\bf 15} (1998) 799
  [arXiv:gr-qc/9708025].

\bibitem{gambini}
  R.~Gambini and J.~Pullin,
  ``Consistent discretizations as a road to quantum gravity,''
  arXiv:gr-qc/0512065.\\
  R.~Gambini and J.~Pullin,
  ``Consistent discretization and canonical classical and quantum Regge
  calculus,''
  Int.\ J.\ Mod.\ Phys.\  D {\bf 15}, 1699 (2006)
  [arXiv:gr-qc/0511096].

\bibitem{alex}
  A.~Perez,
  ``On the regularization ambiguities in loop quantum gravity,''
  Phys.\ Rev.\  D {\bf 73} (2006) 044007
  [arXiv:gr-qc/0509118].




\bibitem{galassi}
  M.~Galassi,
  ``Lapse and shift in Regge calculus,''
  Phys.\ Rev.\  D {\bf 47} (1993) 3254.




\bibitem{hamber}
  H.~W.~Hamber and R.~M.~Williams,
  ``Gauge invariance in simplicial gravity,''
  Nucl.\ Phys.\  B {\bf 487} (1997) 345
  [arXiv:hep-th/9607153].

\bibitem{david}
  L.~Freidel and D.~Louapre,
  ``Diffeomorphisms and spin foam models,''
  Nucl.\ Phys.\  B {\bf 662} (2003) 279
  [arXiv:gr-qc/0212001].

\bibitem{wael} 
H.~Waelbroeck and J.~A.~Zapata,
  ``A Hamiltonian formulation of topological gravity,''
  Class.\ Quant.\ Grav.\  {\bf 11}, 989 (1994)
  [arXiv:gr-qc/9311035].

\bibitem{zap}
  J.~A.~Zapata,
  ``Topological Lattice Gravity Using Self-Dual Variables,''
  Class.\ Quant.\ Grav.\  {\bf 13}, 2617 (1996)
  [arXiv:gr-qc/9603030].


\bibitem{bdlf3d}
  B.~Dittrich, L.~Freidel and S.~Speziale,
  ``Linearized dynamics from the 4-simplex Regge action,''
  Phys.\ Rev.\  D {\bf 76} (2007) 104020
  [arXiv:0707.4513 [gr-qc]].


\bibitem{bdbb}
  B.~Bahr and B.~Dittrich,
  ``Breaking and restoring of diffeomorphism symmetry in discrete gravity,''
  arXiv:0909.5688 [gr-qc].
\\
  B.~Bahr and B.~Dittrich,
  ``Regge calculus from a new angle,''
  New J.\ Phys.\  {\bf 12} (2010) 033010
  [arXiv:0907.4325 [gr-qc]].
\\
  B.~Bahr and B.~Dittrich,
  ``Improved and Perfect Actions in Discrete Gravity,''
  Phys.\ Rev.\  D {\bf 80} (2009) 124030
  [arXiv:0907.4323 [gr-qc]].





\bibitem{pleb}
J.~F.~Plebanski,
  ``On the separation of Einsteinian substructures,''
  J.\ Math.\ Phys.\  {\bf 18}, 2511 (1977).

\bibitem{tregge}
T.~Regge,
  ``General Relativity without coordinates''
  Nuovo Cim.\  {\bf 19}, 558 (1961).





\bibitem{barrettarea}
J.~W.~Barrett, M.~Rocek and R.~M.~Williams,
  ``A note on area variables in Regge calculus,''
  Class.\ Quant.\ Grav.\  {\bf 16}, 1373 (1999)
  [arXiv:gr-qc/9710056].\\
J.~Makela and R.~M.~Williams,
  ``Constraints on area variables in Regge calculus,''
  Class.\ Quant.\ Grav.\  {\bf 18}, L43 (2001)
  [arXiv:gr-qc/0011006].\\
 C.~Wainwright and R.~M.~Williams,
  ``Area Regge calculus and discontinuous metrics,''
  Class.\ Quant.\ Grav.\  {\bf 21} (2004) 4865
  [arXiv:gr-qc/0405031].

\bibitem{immirzi}
  G.~Immirzi,
  ``Quantizing Regge calculus,''
  Class.\ Quant.\ Grav.\  {\bf 13} (1996) 2385
  [arXiv:gr-qc/9512040].\\
  G.~Immirzi,
  ``Regge calculus and Ashtekar variables,''
  Class.\ Quant.\ Grav.\  {\bf 11} (1994) 1971
  [arXiv:gr-qc/9402004].


\bibitem{makela}
  J.~Makela,
  ``On the phase space coordinates and the Hamiltonian constraint of Regge calculus,''
  Phys.\ Rev.\  D {\bf 49} (1994) 2882.

\bibitem{world}
M.~P.~Reisenberger,
  ``A lattice worldsheet sum for 4-d Euclidean general relativity,''
  arXiv:gr-qc/9711052.\\  
 R.~De Pietri and L.~Freidel,
  ``so(4) Plebanski Action and Relativistic Spin Foam Model,''
  Class.\ Quant.\ Grav.\  {\bf 16}, 2187 (1999)
  [arXiv:gr-qc/9804071].\\
M.~P.~Reisenberger and C.~Rovelli,
  ``Spacetime as a Feynman diagram: The connection formulation,''
  Class.\ Quant.\ Grav.\  {\bf 18}, 121 (2001)
  [arXiv:gr-qc/0002095].\\ 
A.~Perez,
  ``Spin foam models for quantum gravity,''
  Class.\ Quant.\ Grav.\  {\bf 20} (2003) R43
  [arXiv:gr-qc/0301113].


\bibitem{epr}
J.~Engle, R.~Pereira and C.~Rovelli,
  ``The loop-quantum-gravity vertex-amplitude,''
  Phys.\ Rev.\ Lett.\  {\bf 99}, 161301 (2007)
  [arXiv:0705.2388 [gr-qc]].\\
 J.~Engle, R.~Pereira and C.~Rovelli,
  ``Flipped spinfoam vertex and loop gravity,''
  Nucl.\ Phys.\  B {\bf 798}, 251 (2008)
  [arXiv:0708.1236 [gr-qc]].


\bibitem{simet}
  E.~R.~Livine and S.~Speziale,
  ``A new spinfoam vertex for quantum gravity,''
  Phys.\ Rev.\  D {\bf 76} (2007) 084028
  [arXiv:0705.0674 [gr-qc]].\\
  E.~R.~Livine and S.~Speziale,
  ``Consistently Solving the Simplicity Constraints for Spinfoam Quantum Gravity,''
  arXiv:0708.1915 [gr-qc].


\bibitem{fk}
L.~Freidel and K.~Krasnov,
  ``A New Spin Foam Model for 4d Gravity,''
  Class.\ Quant.\ Grav.\  {\bf 25}, 125018 (2008)
  [arXiv:0708.1595 [gr-qc]].

\bibitem{alexandrov}
  S.~Alexandrov,
  ``Simplicity and closure constraints in spin foam models of gravity,''
  arXiv:0802.3389 [gr-qc].

\bibitem{monte} 
  M.~Montesinos,
  ``Alternative symplectic structures for SO(3,1) and SO(4)  four-dimensional
  BF theories,''
  Class.\ Quant.\ Grav.\  {\bf 23}, 2267 (2006)
  [arXiv:gr-qc/0603076].


\bibitem{canpleb}
  E.~Buffenoir, M.~Henneaux, K.~Noui and Ph.~Roche,
  ``Hamiltonian analysis of Plebanski theory,''
  Class.\ Quant.\ Grav.\  {\bf 21} (2004) 5203
  [arXiv:gr-qc/0404041].\\
  S.~Alexandrov, E.~Buffenoir and P.~Roche,
  ``Plebanski theory and covariant canonical formulation,''
  Class.\ Quant.\ Grav.\  {\bf 24} (2007) 2809
  [arXiv:gr-qc/0612071].


\bibitem{bdlf} B.~Dittrich, L.~Freidel, to appear


\bibitem{reisenberger1}
  J.~C.~Baez,
  Class.\ Quant.\ Grav.\  {\bf 15} (1998) 1827
  [arXiv:gr-qc/9709052].\\
  M.~P.~Reisenberger,
  ``Classical Euclidean general relativity from *left-handed area =
  right-handed area*,''
  arXiv:gr-qc/9804061.


\bibitem{puzio}
 L.~Freidel, K.~Krasnov and R.~Puzio,
  ``BF description of higher-dimensional gravity theories,''
  Adv.\ Theor.\ Math.\ Phys.\  {\bf 3}, 1289 (1999)
  [arXiv:hep-th/9901069].\\
J.~C.~Baez and J.~W.~Barrett,
  ``The quantum tetrahedron in 3 and 4 dimensions,''
  Adv.\ Theor.\ Math.\ Phys.\  {\bf 3}, 815 (1999)
  [arXiv:gr-qc/9903060].

\bibitem{jan} 
 J.~Ambjorn, B.~Durhuus and T.~Jonsson,
  ``Quantum geometry. A statistical field theory approach,''
{\it  Cambridge, UK: Univ. Pr., 1997. (Cambridge Monographs in Mathematical Physics). }

\bibitem{benni} B.~Bahr and B.~Dittrich,
  ``(Broken) Gauge Symmetries and Constraints in Regge Calculus,''
  Class.\ Quant.\ Grav.\  {\bf 26} (2009) 225011
  [arXiv:0905.1670 [gr-qc]].

\bibitem{gft}
  L.~Freidel,
  ``Group field theory: An overview,''
  Int.\ J.\ Theor.\ Phys.\  {\bf 44} (2005) 1769
  [arXiv:hep-th/0505016].

\bibitem{oeckl}
  R.~Oeckl,
  ``A 'general boundary' formulation for quantum mechanics and quantum
  gravity,''
  Phys.\ Lett.\  B {\bf 575}, 318 (2003)
  [arXiv:hep-th/0306025].
 
\bibitem{aristide}
  A.~Baratin and L.~Freidel,
  ``Hidden quantum gravity in 4d Feynman diagrams: Emergence of spin foams,''
  Class.\ Quant.\ Grav.\  {\bf 24} (2007) 2027
  [arXiv:hep-th/0611042].
 
\bibitem{tent}
  J.~W.~Barrett, M.~Galassi, W.~A.~Miller, R.~D.~Sorkin, P.~A.~Tuckey and R.~M.~Williams,
  ``A Paralellizable implicit evolution scheme for Regge calculus,''
  Int.\ J.\ Theor.\ Phys.\  {\bf 36} (1997) 815
  [arXiv:gr-qc/9411008].


\bibitem{thiemann}
A.~Ashtekar, J.~Lewandowski, D.~Marolf, J.~Mourao and T.~Thiemann,
  ``Quantization of diffeomorphism invariant theories of connections with local
  degrees of freedom,''
  J.\ Math.\ Phys.\  {\bf 36}, 6456 (1995)
  [arXiv:gr-qc/9504018].

\bibitem{bianchi}
 E.~Bianchi,
  ``The length operator in Loop Quantum Gravity,''
  arXiv:0806.4710 [gr-qc].


\bibitem{winkler}
  T.~Thiemann,
  ``Gauge field theory coherent states (GCS). I: General properties,''
  Class.\ Quant.\ Grav.\  {\bf 18} (2001) 2025
  [arXiv:hep-th/0005233].\\
 T.~Thiemann and O.~Winkler,
  ``Gauge field theory coherent states (GCS). II: Peakedness properties,''
  Class.\ Quant.\ Grav.\  {\bf 18}, 2561 (2001)
  [arXiv:hep-th/0005237].



\bibitem{aqg}
  K.~Giesel and T.~Thiemann,
  ``Algebraic quantum gravity (AQG). I: Conceptual setup,''
  Class.\ Quant.\ Grav.\  {\bf 24} (2007) 2465
  [arXiv:gr-qc/0607099].


\bibitem{carl}
C.~Rovelli,
  ``Graviton propagator from background-independent quantum gravity,''
  Phys.\ Rev.\ Lett.\  {\bf 97} (2006) 151301
  [arXiv:gr-qc/0508124].\\
E.~Bianchi, L.~Modesto, C.~Rovelli and S.~Speziale,
  ``Graviton propagator in loop quantum gravity,''
    Class.\ Quant.\ Grav.\  {\bf 23} (2006) 6989
  [arXiv:gr-qc/0604044].\\
E.~Alesci and C.~Rovelli,
  ``The complete LQG propagator: I. Difficulties with the Barrett-Crane
  Phys.\ Rev.\  D {\bf 76} (2007) 104012
  [arXiv:0708.0883 [gr-qc]].\\
  E.~Alesci and C.~Rovelli,
  ``The complete LQG propagator: II. Asymptotic behavior of the vertex,''
  Phys.\ Rev.\  D {\bf 77}, 044024 (2008)
  [arXiv:0711.1284 [gr-qc]].


\bibitem{bdry}
  R.~Pereira,
  ``Lorentzian LQG vertex amplitude,''
  Class.\ Quant.\ Grav.\  {\bf 25} (2008) 085013
  [arXiv:0710.5043 [gr-qc]].\\
 J.~Engle, E.~Livine, R.~Pereira and C.~Rovelli,
  ``LQG vertex with finite Immirzi parameter,''
  Nucl.\ Phys.\  B {\bf 799}, 136 (2008)
  [arXiv:0711.0146 [gr-qc]].\\
F.~Conrady and L.~Freidel,
  ``Path integral representation of spin foam models of 4d gravity,''
  arXiv:0806.4640 [gr-qc].


\bibitem{lf} L. Freidel, private communication


\bibitem{rovcos}
C.~Rovelli and F.~Vidotto,
  ``Stepping out of Homogeneity in Loop Quantum Cosmology,''
  arXiv:0805.4585 [gr-qc].



\bibitem{tt}
T.~Thiemann,
  ``Quantum spin dynamics (QSD). VII: Symplectic structures and continuum
  lattice formulations of gauge field theories,''
  Class.\ Quant.\ Grav.\  {\bf 18} (2001) 3293
  [arXiv:hep-th/0005232].






\end{thebibliography}
\end{document}